\documentclass[a4paper,fleqn,usenatbib]{mnras}

\usepackage[T1]{fontenc}
\usepackage{ae,aecompl}

\usepackage{graphicx}	
\usepackage{amsmath}	
\usepackage{amssymb}	

\newcommand{\kms}{~km~s$^{-1}$}

\title[Water Fountains and Circumstellar Envelopes]{
Do Water Fountain Jets Really Indicate the Onset of the Morphological 
Metamorphosis of Circumstellar Envelopes?
}

\author[B. H. K. Yung et al.]{
Bosco H. K. Yung,$^{1,2}$\thanks{E-mail: byung@ncac.torun.pl (BHKY)}
Jun-ichi Nakashima,$^{3,2}$
Chih-Hao Hsia$^{5,4,2}$
and Hiroshi Imai$^{6}$
\\
$^{1}$Nicolaus Copernicus Astronomical Center, Rabia\'nka 8, 87-100 Toru\'n, Poland\\
$^{2}$Department of Physics, The University of Hong Kong, Pokfulam Road, Hong Kong\\
$^{3}$Department of Astronomy and Geodesy, Ural Federal University, Lenin Ave. 51, 620000, Ekaterinburg, Russia\\
$^{4}$Laboratory for Space Research, Faculty of Science, The University of Hong Kong, Pokfulam, Hong Kong\\
$^{5}$Space Science Insitute, Macau University of Science and Technology, Avenida Wai Long, Taipa, Macau\\
$^{6}$Graduate School of Science and Engineering, Kagoshima University, 1-21-35 Korimoto, 890-0065 Kagoshima, Japan
}

\date{Accepted XXX. Received YYY; in original form ZZZ}

\pubyear{2016}

\begin{document}
\label{firstpage}
\pagerange{\pageref{firstpage}--\pageref{lastpage}}
\maketitle

\begin{abstract}
The small-scale bipolar jets having short dynamical ages 
from ``water fountain (WF)'' sources are regarded as an indication of the 
onset of circumstellar envelope morphological metamorphosis of
intermediate-mass stars. Such process
usually happens at the end of the asymptotic giant branch (AGB) phase. 
However, recent studies found that WFs could be AGB stars or even early
planetary nebulae. 
This fact prompted the idea that WFs may not necessarily be
objects at the beginning of the morphological transition process.
In the present work, we show that WFs could have different envelope 
morphologies by studying their spectral energy 
distribution profiles. Some WFs have spherical envelopes that 
resembles usual AGB stars, while others have aspherical envelopes which
are more common to post-AGB stars.
The results imply that WFs may not represent the earliest stage of the
morphological metamorphosis. We further argue that the dynamical age of a WF
jet, which can be calculated from maser proper motions, may not be the real 
age of the jet. The dynamical age cannot be used to 
justify the moment when the envelope begins to 
become aspherical, nor to tell the concrete evolutionary status of the 
object. A WF jet could be the innermost part of a larger well-developed jet,
which is not necessarily a young jet. 
\end{abstract}

\begin{keywords}
infrared: stars -- radiative transfer -- stars: AGB and post-AGB -- 
stars: evolution -- stars: mass-loss -- stars: winds, outflows
\end{keywords}



\section{Introduction}
\label{sec:intro}

Planetary nebulae~(PNe) that evolved from stars with masses about 
1--8~$M_{\odot}$ have different morphologies, such as spherical, 
bipolar, or even multi-polar \citep[see,][for a review]{kwok10pasa}. On the 
contrary, their 
progenitors, the mass-losing asymptotic giant branch~(AGB) stars, are mostly 
just spherical in terms of their circumstellar envelopes (or envelopes). 
A vast change in morphology is expected to have happened in
between the AGB and PN phases, which is called the post-AGB phase. 
It is suggested that high velocity jets emerged from 
late/post-AGB stars play a key role in shaping PNe 
\citep{sahai98aj,sahai15apjl}. 
Nonetheless, the exact jet formation mechanism and the jet-envelope 
interaction process are still unclear. While larger bipolar structures 
likely resulting from jets can be observed in infrared or optical images 
\citep[e.g.,][]{lagadec11mnras,siodmiak08apj,sahai07aj}, some jets can only 
be revealed by interferometric observation of molecular lines such as CO. 
The spatial extent of such molecular jets could 
reach to a few thousands AU from the star, e.g., $\sim$6000~AU for the
post-AGB star IRAS~08005$-$2356 \citep{sahai15apjl}, with velocity 
$>$100\kms. Some molecular outflows (jets or torii)
were also observed occasionally in AGB stars, 
e.g., V~Hydrae \citep{hirano04apjl} and X~Herculis \citep{nakashima05apj}.

In the case of oxygen-rich stars, which are evolved stars with more oxygen
than carbon in the envelopes, there are a type of objects
called the ``water fountains~(WFs)'' that have relatively ``tiny''
collimated jets traced by 22~GHz H$_{2}$O maser emission 
\citep[see,][for reviews on WFs]{imai07iaup,desmurs12iaup}. 
The spatial extent of WF jets is usually relatively small, which is
of order $10^{2}$--$10^{3}$~AU 
\citep[e.g.,][]{imai02nature,boboltz07apj,day10apj,yung11apj}.
These jets are characterized by the large spectral velocity coverage 
(usually $>$50\kms) of 
H$_{2}$O maser emission, which exceeds the usual 1612~MHz OH maser coverage 
\citep[$\leq$25\kms,][]{hekkert89aas}.
WFs are relatively rare as there are only 16 confirmed examples known to date,
plus a few candidates reported in 
\citet{yung13apj,yung14apj} and \citet{gomez15aa}.
There are several possible reasons for why they are rare. 
\citet{gomez15aa} suggests that the WF phenomenon may in fact not 
that uncommon, but the maser emission from most of those objects are too weak
to be detected; the exact mechanism is unknown but it could be related to the
masses of the star progenitors. Another reason is related to the viewing angle
to the jet as explored in \citet{yung13apj}: even if the jet velocity is high,
when the jet orientation is rather edge-on, the H$_{2}$O maser velocity
coverage may still be smaller than that of the corresponding OH maser due to
projection. Chemical bias could also be a reason because we are focusing only
on Oxygen-rich stars, and there is currently no way to observe WF-equivalence
for carbon-rich stars.

The dynamical ages of WF jets are found to be very short 
\citep[$\leq$100~years,][]{imai07iaup}. Together with the small jet sizes,
most of the WFs are thought to be objects at the beginning stage of the
morphological transition, which usually happens at the early 
post-AGB phase \citep[e.g,][]{suarez08apj,walsh09mnras,day10apj,desmurs12iaup}. 
However, even though a majority of the WFs are very likely to be post-AGB stars, 
e.g., IRAS~16342$-$3814 \citep{sahai99apj}, IRAS~18113$-$2503 
\citep{gomez11apj}, IRAS~18286$-$0959 \citep{yung11apj}, and 
IRAS~18455$+$0448 \citep{vlemmings14aa},
there are some clear exceptions. W~43A shows flux variation in OH maser with
a period of 400 days, which is caused by the periodic variation in infrared 
emission due to envelope pulsation \citep{herman85aas,imai02nature}. 
SiO maser is also detected toward this object 
\citep{nakashima03pasj2}. 
These indicates that W~43A might be still in the AGB phase. 
On the other hand,
IRAS~15103$-$5754 is suggested to be a PN candidate but it also exhibits
WF characteristics (i.e. high velocity H$_{2}$O maser emission). The
detection of Ne~II emission line and free-free continuum emission give
evidence to the PN status, making this object 
the first WF-PN ever found \citep{suarez09aa,lagadec11mnras,gomez15apj}. 
Given the above exceptions, it is suspected that the WF-type objects 
are not necessarily in the short early post-AGB phase or transitional phase.
They may also not representing the onset of the 
morphological metamorphosis of the AGB envelopes. Furthermore,
the relationship between the WF maser jets, the larger-scale molecular jets,
and the ultimate large bipolar feature visible in infrared, is not known.

To find out the role of WF jets in such morphological changing process, one 
way is to examine whether there is a direct correlation between the WF jets 
and the 
envelope morphology of the WF sources. This can be done by looking
at the infrared spectral energy distributions~(SEDs). It is because the 
SED profile can give constraints to the possible 
envelope morphology of an AGB/post-AGB star.
In this paper, SEDs of the known WFs are presented together with dust radiative 
transfer models. However, the main goal here is to explore whether the dust 
envelopes of WFs have departed from spherical symmetry, but not the 
deep interpretation of the model parameters.
Some of the WFs are shown to have aspherical structures
under high resolution infrared images 
\citep[e.g.,][]{lagadec11mnras,ramos12aa},
nonetheless a number of them may still have spherical envelopes.
This work is also the first attempt to study WF envelopes collectively
by radiative transfer models. This simple but effective approach will be 
useful in the future for statistical studies of stellar maser sources, which 
will be detected with
new telescopes such as the \textit{Square Kilometre Array} (\textit{SKA})
and the \textit{Five hundred metre Aperture Spherical Telescope} 
(\textit{FAST}). The data size resulted from these anticipated maser surveys 
will be huge, and hence developing quick analysis approaches is very 
meaningful.

The data and SED analysis are described in 
Section~\ref{sec:ana}. The results and interpretations are
given in Section~\ref{sec:res}, followed by the discussion 
in Section~\ref{sec:dis}.
Finally, the 
conclusions are presented in Section~\ref{sec:con}.

\section{SED Analysis}
\label{sec:ana}

\subsection{Method}
\label{ssec:met}

PNe are visible in optical and
hence morphological classification can be made from optical images
\citep[e.g.,][]{ueta00apj}. On the contrary, AGB and some post-AGB stars
are optically opaque due to their thick envelopes, and the detailed envelope
structures are mostly visible in infrared wavelengths. Therefore, in this 
study we focus on the infrared SED data.
The main concern here is whether WFs have spherical or aspherical 
(infrared) envelopes, hence an one-dimensional radiative transfer code is 
chosen for this analysis. However this is not a very ``standard'' radiative
transfer analysis. In our case the obtained parameters are not of our
greatest interest, instead we focus on whether the models are ``fit'' or 
``unfit'' to the SED profiles.

The SED of an AGB/post-AGB star with clear aspherical envelope is unlikely to 
be reproduced by one-dimensional models. However, to confirm whether there is
really no good fits for an SED is not so straightforward. 
It is because the SED profile shape is affected not only by the envelope 
morphology, but also by the stellar temperature, chemistry and the 
number of dust components, etc. In order to be sure that a certain SED cannot 
be fit, 
in principle we have to explore all physically possible combinations of 
the parameters. Nonetheless this is practically not feasible, therefore 
specific cases would be examined to help excluding some less sensitive
parameters 
(more in Section~\ref{ssec:rad}). After that we can be more confident that
the ``unfit'' SEDs are associated with aspherical envelope morphology, but
not due to other parameters.  
If WFs could have both spherical and aspherical envelopes, then it means that
not all of them are at the same (early) stage of the morphology changing
process. 

To justify the effectiveness of this fitting method, six standard 
AGB stars and six characteristic post-AGB stars were also included in
our sample as control for making comparison (more in Section~\ref{ssec:jus}). 
Majority of the AGB stars are 
expected to have spherical envelopes, but there are also a few clear 
exceptions due to the existence of (early) jets, such as
V~Hydrae \citep{hirano04apjl}, X~Herculis \citep{nakashima05apj}, and 
CIT~6 \citep[e.g.,][]{monnier00apj}.
Aspherical features are more commonly found in  
post-AGB stars. It is expected that good SED fits can be obtained from the
one-dimensional code for most of the spherical AGB stars, but not always 
for the post-AGB stars
(more in Section~\ref{ssec:jus}).
We believe that if some
meticulous manipulation is done on the parametric values it might be 
possible to fit SEDs of aspherical objects even with one-dimensional models,
but this does not mean the solution is really physical. We try to avoid
such over-manipulation by limiting our fitting using the three defining 
parameters.

\subsection{Data Retrieved}
\label{ssec:dat}

The research targets include 17 objects. They are either known WFs to which
the existence of bipolar jets has been confirmed by interferometric 
observations, or WF candidates with possible AGB/post-AGB evolutionary 
status that show WF spectral characteristic in single-dish observations 
(i.e. velocity coverage of H$_{2}$O maser is larger than that of OH maser). 
In this paper these objects are all treated as ``WFs''. 
The object list with corresponding references is given in 
Table~\ref{tab:objs}. 

The infrared photometric data (from 1.25 to 160~$\mu$m) used to construct 
the SEDs were collected from the point source catalogues of 
\textit{Two Micron All Sky Survey} 
\citep[\textit{2MASS},][]{skrutskie06aj}, 
\textit{Wide-field Infrared Survey Explorer} 
\citep[\textit{WISE},][]{wright10aj},
\textit{Infrared Astronomical Satellite}
\citep[\textit{IRAS},][]{neugebauer84apjl},
\textit{Midcourse Space Experiment} 
\citep[\textit{MSX},][]{egan03msx}, and
\textit{AKARI} \citep{kataza10akari,yamamura10akari}. 
More data were obtained from the images taken by the
\textit{Infrared Array Camera} \citep[\textit{IRAC},][]{fazio04apjs} and the
\textit{Multiband Imaging Photometer for Spitzer} 
\citep[\textit{MIPS},][]{rieke04apjs} mounted on the 
\textit{Spitzer Space Telescope}. 
The way of doing photometry on these images is described in 
\citet{hsia14aa}. All the above data are presented in 
Appendix~\ref{app:data}. 

For W~43A, we have additional sub-millimetre flux data obtained by the 
\textit{Very Large Array}~(\textit{VLA}): 4.02~mJy at 7~mm \citep{imai05apj}, 
new observation results from the
\textit{Berkeley Illinois Maryland Association}~(\textit{BIMA}) 
\textit{Millimeter Array},
\textit{Nobeyama Millimetre Array} 
(\textit{NMA}), and the \textit{Jansky-VLA}~(\textit{JVLA}, upgraded
version of the original \textit{VLA}), which covers the wavelengths from about 
1.3~mm to 30~mm. 
The \textit{BIMA} observations (project code: t817d229) were conducted on
7 and 12 September 2003 with the D-array. Uranus and MWC~349 were used as
flux calibrators, and 1751$+$096 as phase calibrator. 
Calibration and image synthesis were done with MIRIAD
\footnote{https://bima.astro.umd.edu/miriad/}.
The flux obtained was
250~mJy at 1.3~mm with root-mean-square~(rms) noise about 26.8~mJy per beam.
The \textit{NMA} observations were carried out on 25--26 
December 2002 with
D-configuration, 10--11 January 2003 with AB-configuration, and 
27--28 March 2003 with C-configuration.
The \textit{JVLA} observations (project code: 13A-041) were carried out on 
9 June 2013 with Q-band (46.0~GHz), 12 June 2013 with X-band (10.1~GHz), 
and 13 June 2013 with K-band (24.2~GHz). J184603.8$-$000338 was used as 
the phase calibrator for X- and K-bands, and J185146.7$+$003532 for 
Q-band. OT081 was used as the bandpass and flux calibrators for 
K- and Q-bands, and 3C286 for X-band.
Calibration and image synthesis of the \textit{NMA} data were done with 
AIPS\footnote{http://www.aips.nrao.edu/index.shtml}, 
while for the \textit{JVLA} data, CASA\footnote{https://casa.nrao.edu/index.shtml} 
was used in producing image cubes and then analysed
with AIPS. Details of the NMA and JVLA observation 
results are given in Table~\ref{tab:submm}. 

We understand
that there may exist other photometric/spectral data for some of the objects,
however, the above data are sufficient to construct unambiguous SEDs, which
is good enough for our scientific purpose. Our focus is on the general
profile shape of the SEDs, but not the detailed chemistry of the line 
features.
The effect of interstellar extinction are known to be significant 
especially toward the direction near to the 
Galactic plane and bulge. Such effect is prominent for shorter
wavelengths.
Thus, for data with wavelengths shorter than 8~$\mu$m,
correction on interstellar extinction is necessary. The method used is
described in \citet{howarth83mnras}, and the required extinction coefficient
for each object were determined from the Galactic reddening maps given by 
\citet{schlegel98apj} and \citet{schlafly11apj}. The extinction values 
$A({\rm V})$ are listed in Table~\ref{tab:2mass}.

\begin{table*}
\caption{List of objects studied. See text for the details of the DUSTY fit. 
Envelope morphological information obtained from images, if any, is given in 
the ``Image'' column.  
For the water fountains and the candidates, 
the representative papers discussing the maser kinematics and evolutionary
statuses are given in the last column.}
\label{tab:objs}
\begin{tabular}{llllll}
\hline
{Object} &
{R.A.} &
{Decl.} &
{Image}$^{\rm a}$ &
{DUSTY$^{\rm b}$} &
{WF Ref.} \\

&
{(J2000.0)} &
{(J2000.0)} &
&
&
\\
\hline

\multicolumn{6}{c}{Water Fountains} \\
\hline
IRAS~15445$-$5449 & 15 48 19.37 & $-$54 58 21.2 & As & $\cdots$ & \citet{sanchez11mnras}  \\
IRAS~15544$-$5332 & 15 58 18.40 & $-$53 40 40.0 & $\cdots$ & $\cdots$ & \citet{deacon07apj} \\
IRAS~16342$-$3814 & 16 37 39.91 & $-$38 20 17.3 & As & ND & \citet{sahai99apj} \\
IRAS~16552$-$3050 & 16 58 27.80 & $-$30 55 06.2 & $\cdots$ & NE & \citet{suarez08apj} \\
IRAS~18043$-$2116 & 18 07 21.10 & $-$21 16 14.2 & Un & $\cdots$ & \citet{walsh09mnras} \\
IRAS~18056$-$1514 & 18 08 28.40 & $-$15 13 30.0 & $\cdots$ & ND & \citet{yung13apj} \\
IRAS~18113$-$2503 & 18 14 27.26 & $-$25 03 00.4 & $\cdots$ & D? & \citet{gomez11apj} \\
OH~12.8$-$0.9 & 18 16 49.23 & $-$18 15 01.8 & Un & S & \citet{boboltz07apj} \\
IRAS~18286$-$0959 & 18 31 22.93 & $-$09 57 21.7 & Un & $\cdots$ & \citet{yung11apj} \\
OH~16.3$-$3.0 & 18 31 31.51 & $-$16 08 46.5 & $\cdots$ & NE & \citet{yung14apj} \\
W~43A & 18 47 41.16 & $-$01 45 11.5 & As & ND & \citet{imai02nature} \\
IRAS~18455$+$0448 & 18 48 02.30 & $+$04 51 30.5 & $\cdots$ & S & \citet{vlemmings14aa} \\
IRAS~18460$-$0151 & 18 48 42.80 & $-$01 48 40.0 & Un & ND & \citet{imai13apj2} \\
IRAS~18596$+$0315 & 19 02 06.28 & $+$03 20 16.3 & $\cdots$ & NE & \citet{amiri11aa} \\
IRAS~19134$+$2131 & 19 15 35.22 & $+$21 36 33.9 & Un & ND & \citet{imai07apj} \\
IRAS~19190$+$1102 & 19 21 25.09 & $+$11 08 41.0 & $\cdots$ & ND & \citet{day10apj} \\
IRAS~19356$+$0754 & 19 38 01.90 & $+$08 01 32.0 & $\cdots$ & D & \citet{yung14apj} \\
\hline
\multicolumn{6}{c}{ Control AGB Stars} \\
\hline
IRAS~14247$+$0454 & 14 27 16.39 & $+$04 40 41.1 & $\cdots$ & S & $\cdots$ \\
IRAS~18556$+$0811 & 18 58 04.23 & $+$08 15 30.8 & $\cdots$ & S & $\cdots$ \\
IRAS~19149$+$1638 & 19 17 11.55 & $+$16 43 54.5 & $\cdots$ & S & $\cdots$ \\
IRAS~19312$+$1130 & 19 33 34.56 & $+$11 37 02.6 & $\cdots$ & S & $\cdots$ \\
IRAS~19395$+$1827 & 19 41 44.55 & $+$18 34 25.8 & $\cdots$ & S & $\cdots$ \\
IRAS~19495$+$0835 & 19 51 57.71 & $+$08 42 54.6 & $\cdots$ & S & $\cdots$ \\
\hline
\multicolumn{6}{c}{Control Post-AGB Stars} \\
\hline
IRAS~07134$+$1005 & 07 16 10.26 & $+$09 59 48.0 & As & $\cdots$ & $\cdots$ \\
OH~231.8$+$4.2 & 07 42 16.95 & $-$14 42 50.2 & As & $\cdots$ & $\cdots$ \\
IRAS~17441$-$2441 & 17 47 13.49 & $-$24 12 51.4 & As & ND & $\cdots$ \\
IRAS~17534$+$2603 & 17 55 25.19 & $+$26 02 60.0 & Un & S? & $\cdots$ \\
IRAS~20547$+$0247 & 20 57 16.28 & $+$02 58 44.6 & Un & S? & $\cdots$ \\
IRAS22272$+$5435 & 22 29 10.37 & $+$54 51 06.4 & As & D & $\cdots$ \\
\hline
\end{tabular}
\\
\flushleft
$^{\rm a}${Images are taken from \citet{lagadec11mnras} or 
            from the WF reference papers. Envelopes showing extended 
            aspherical
            features are denoted by ``As'', unresolved envelopes are denoted
            by ``Un''.}\\
$^{\rm b}${SED profiles that can be fit by DUSTY models are classified into 
            single-peaked ``S'' or double-peaked ``D''. The extra ``?'' 
            indicates that the SEDs are marginally fit by the models.
            For the profiles that cannot be fit, many of those
            belong to two categories, those with near-infrared
            excess ``NE'' comparing to the models, and those with 
            near-infrared deficit ``ND''. See Section~\ref{ssec:typ} for
            details.}
\end{table*}


\begin{table}
\caption{Ranges and steps of the parameters used in generating the 
DUSTY models.}
\label{tab:para}
\begin{tabular}{lll}

\hline
{Parameter} &
{Range} &
{Step} \\ 
\hline
$T_{\rm eff}$~(K) & 1600--10000 & 200 \\
$T_{\rm d}$~(K)   & 100--1600  & 100 \\
$\tau_{2.2}$      & 0.01--0.1  & 0.005 \\
$\tau_{2.2}$      & 0.11--0.3  & 0.01 \\
$\tau_{2.2}$      & 0.35--0.95 & 0.05 \\
$\tau_{2.2}$      & 1--10      & 1\\
\hline
\end{tabular}

\end{table}

\begin{figure*}
   \includegraphics[scale=0.8]{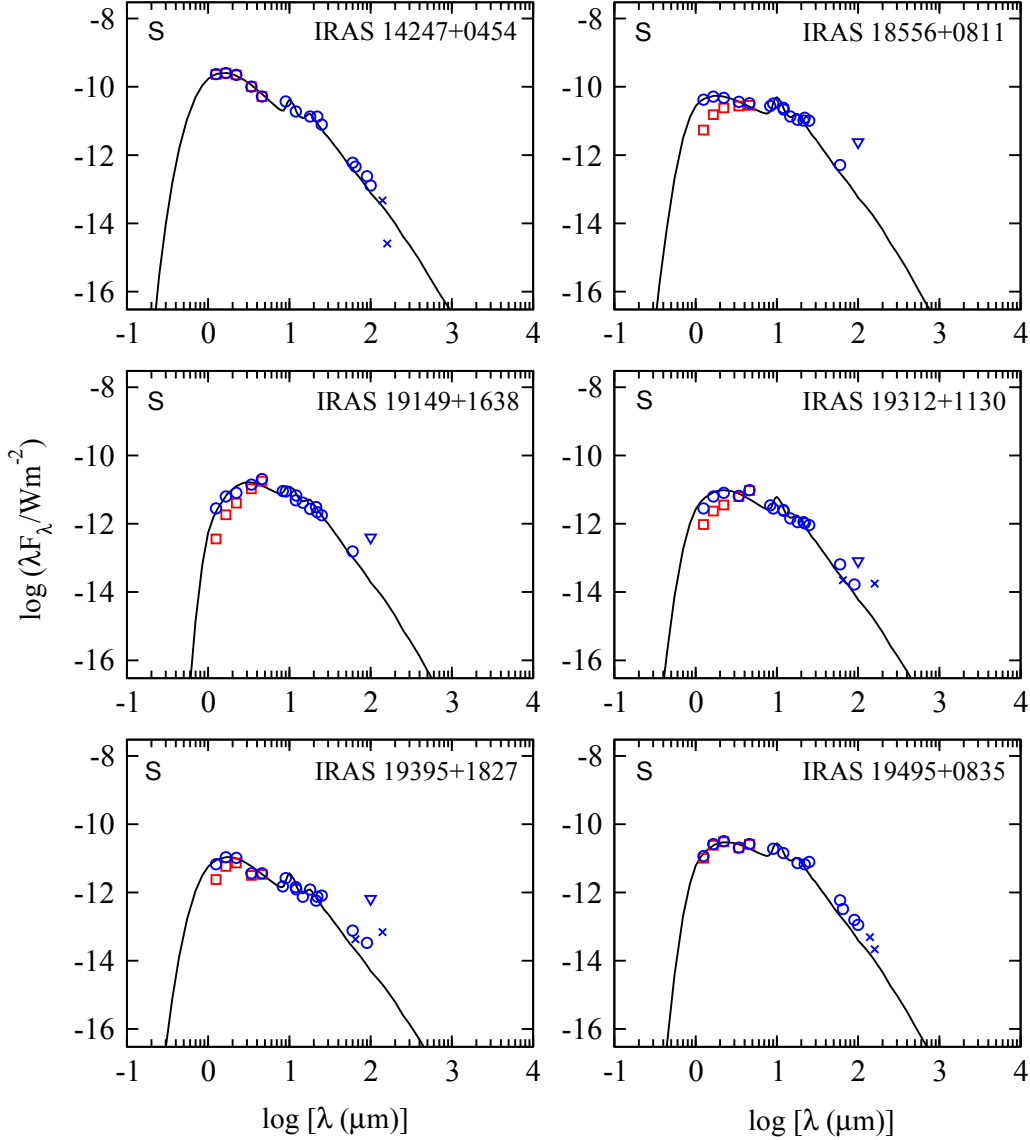}
   \caption{Spectral energy distributions of some standard AGB stars with 
            spherical envelopes.
            The red squares represent original data points without correction 
            on interstellar extinction, while the blue circles represent data 
            points with correction applied. The lower and upper limits of the
            fluxes are indicated by upward and downward triangles, 
            respectively. Flux values with a low-quality flag are indicated by
            crosses. For the triangles and crosses, red colour is used for the
            data points without correction on interstellar extinction, and 
            blue colour is used for the data points with correction applied.
            The black curves represent the best-fit model with the DUSTY 
            code (see text for details). For the cases where good fits are
            available, ``S'' denotes a single-peaked profile, ``D''
            denotes a double-peaked profile, an extra ``?'' means the SED 
            is marginally fit by the model. For the cases without good fits,
            ``NE'' denotes those with near-infrared excess, and
            ``ND'' denotes near-infrared deficit in the profile. 
           }
   \label{fig:sed_agb}
\end{figure*}


\begin{figure*}
   \includegraphics[scale=0.8]{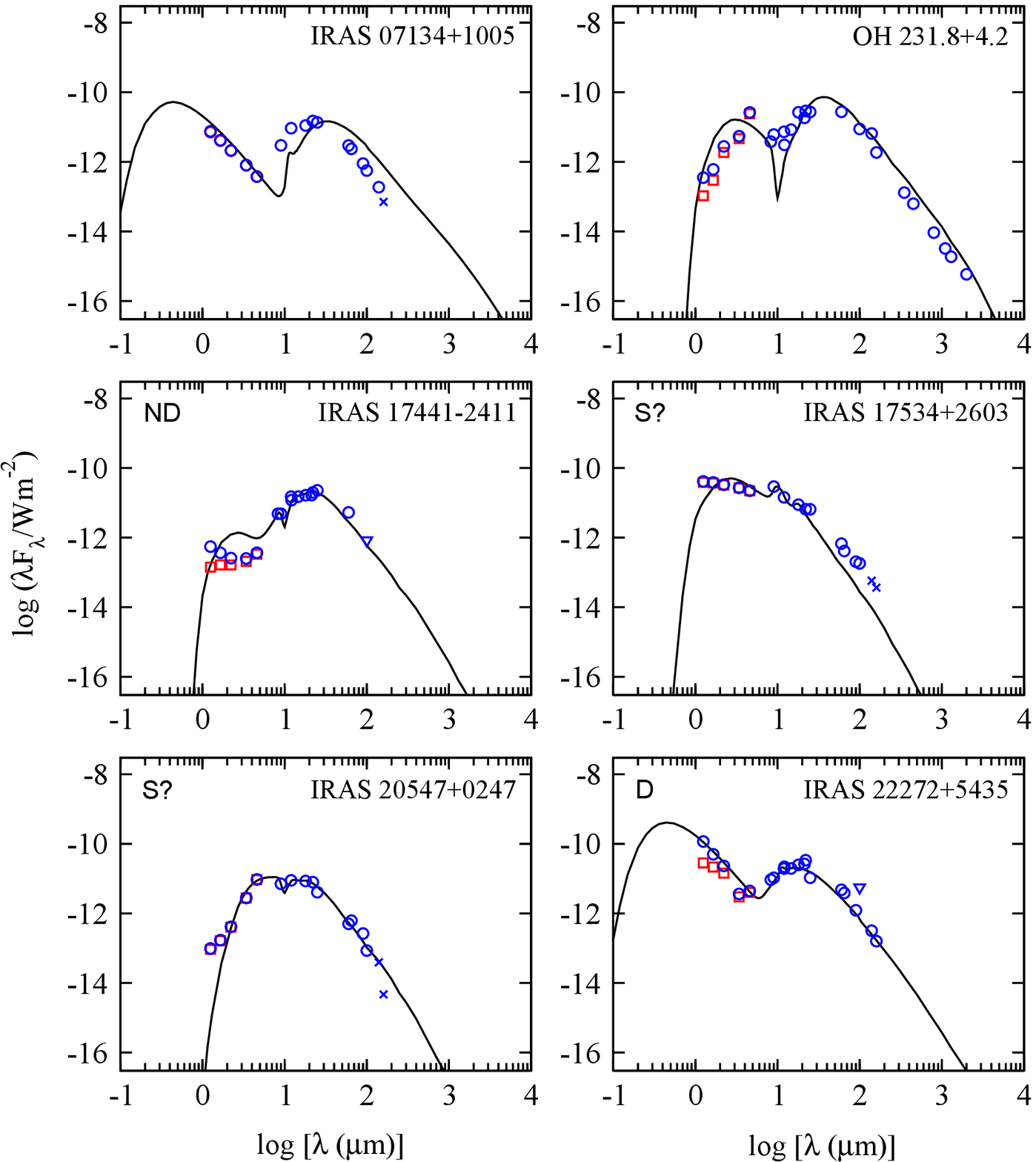}
   \caption{Spectral energy distributions of some post-AGB stars with 
            aspherical structures, for e.g., tori or jets. 
            The notations used in this figure
            are the same as Figure~\ref{fig:sed_agb}.
           }
   \label{fig:sed_ppn}
\end{figure*}


\subsection{One-Dimensional Radiative Transfer Modelling}
\label{ssec:rad}

The one-dimensional dust radiative transfer models that used to fit the SEDs 
were generated by the DUSTY code \citep{ivezic99dusty}. As mentioned before,
even though it is not possible to literally consider all combinations of
different parameters, we tried to justify the most important ones for our
purpose by performing specific tests. At the end we are left with three running
parameters (see below).

We assumed that the only radiation source was coming from a point-source in 
the centre of the spherical envelope. The SEDs from the central
point sources were taken to be Planckian (i.e., black body curves).
We once thought that ultra-violet~(UV) emission was detected from two
of our WFs (IRAS~16552$-$3050 and OH~16.3$-$3.0) when we were searching 
through the source catalogue\footnote{http://galex.stsci.edu/GR6/} of the
\textit{Galaxy Evolution Explorer} (\textit{GALEX}). The UV emission might
be in principle coming from, e.g., a hot binary component such as a white
dwarf, and such binary systems are not rare. However, it is now confirmed 
that the UV detections are not associated
with our two WFs because the separations are too large ($>$0.5\arcmin), and
there exists other sources which are more likely to be the host of the
UV emission. The detected UV fluxes also seem to be unreasonably large 
if we assume a typical case of a binary including a white dwarf. No other
obvious
sign of binary components were observed from our samples according to our
data, hence the single point-source model remains more reasonable. Note that 
we are not excluding binary cases here, but with our current data we do not 
intend to over-interpret this idea. Some possible cases with hot companions
are mentioned though (Sections~\ref{sec:res} and \ref{sec:dis}).

The grain type with 50\% warm (Sil-Wc) and 50\% 
cold (Sil-Oc) silicates was chosen \citep{ossenkopf92aa}. While AGB stars
consist of envelops with warmer dust, the detached envelops of post-AGB stars 
would have colder dust. Since we were uncertain about the exact situation of
each case, we took the 50/50 assumption. In fact, we have tested that 
within our temperature ranges (see Table~\ref{tab:para}), this warm-cold ratio 
has minimal effect to the SED profile shape. More warm dust would shift up the
model curve in mid-infrared by less than a few percent, but still keeping the
curve shape quite precisely. We judge that this small
shift does not affect much because once again our main goal is not to 
determine very accurately the physical conditions of the dust envelopes, but 
to see whether obtaining a reasonable good fit is possible or not. 
The grain size distribution was set to follow the standard 
Mathis-Rumpl-Nordsieck~\cite[MRN,][]{mathis77apj} power law. Regarding the 
analytical profiles, we selected the DUSTY option which supposed the envelope 
expansion was driven by radiation pressure on the dust grains. This was 
suitable for the case of mass-losing stars. The analytic approximation for 
radiatively driven winds was used, where the variation of flux-averaged 
opacity with 
radial distance was assumed to be negligible, so that the hydrodynamics 
equations could be solved analytically. This approximation offered the 
advantage of a much shorter run time, and it was suggested to be suitable for 
most AGB stars \citep{ivezic99dusty}. The envelope thickness was 
assumed to be 10,000 times of its inner radius. We also tested that even
changing this ratio up to 40\% would have a negligible change to our model
curves.

Three running parameters remained for the DUSTY code under the 
above considerations, namely the effective temperature of the
central radiation source ($T_{\rm eff}$), the dust temperature at the inner
envelope boundary ($T_{\rm d}$), and the optical depth at 2.2~$\mu$m 
($\tau_{2.2}$). To find the best fit models for the SEDs, we generated a model 
grid with all combination of the three parameters within defined numerical 
ranges. The setting on
the ranges and steps for each parameter is given in Table~\ref{tab:para}.
The closest fit was obtained by minimizing the sum-of-squares of the flux 
deviations between the observed data points and the model curves (with 
appropriate scaling). Generally speaking, in our cases the good fits are 
those that have the least sum-of-square values $\sim$$10^{-13}$.
Note that in some marginally fit cases, or when multiple fit solutions seem 
plausible, the best fits have to be finally determined qualitatively by eye. 
For instant, we 
have to see if the model can reproduce the key features of an SED line shape,
such as the number of peaks and general changes of slopes. Even when the data 
points are a bit off from the model curve, it could still be considered as a 
good fit if the above general important features are reproduced. These 
cases will be discussed in Section~\ref{sec:res}.

\subsection{Justification with Control Objects}
\label{ssec:jus}

Some control objects were used to test whether the DUSTY models were
sensitive enough to distinguish spherical and aspherical envelopes.
We mainly selected AGB and post-AGB stars with known envelope morphologies 
(e.g., from high resolution infrared images) and reasonably enough 
photometric data points for SED construction. These objects were found in
roughly the same R.A. range of the WFs, as to minimize the effect from any 
possible position-dependent factors, e.g., to avoid huge differences in the 
galactic dust extinction toward the sources. Such extinction will 
affect the SED data in the near-infrared range (Section~\ref{ssec:dat}).

There is usually no high resolution images of typical spherical AGB stars,
therefore we selected such control objects from the SiO maser sources.
Our control AGB stars all exhibit a sharp
single-peaked SiO maser feature \citep{nakashima03pasj2}. This is a good
evidence showing that the corresponding envelopes are spherical, because  
this type of maser spectral profile is produced when the masers are 
tangentially amplified in the spherical envelope.
For the control post-AGB stars,
IRAS~07134$+$1005 is suggested to have a geometrically thick expanding torus
\citep{nakashima09apj}; 
IRAS~22272$+$5435 is so far best
interpreted as a system consists of spherical wind, torus, and a jet
interacting with ambient materials \citep{nakashima12apj}. The 
aspherical structures of
both objects have been revealed via interferometric observations of the
CO line. 
The other four post-AGB stars were selected from the infrared image catalogue
presented in \citet{lagadec11mnras}. Both IRAS~17534$+$2603 and 
IRAS~20547$+$0247 are unresolved objects which could either mean they are
spherical or they are too far away. OH~231.8$+$4.2 and IRAS~17441$-$2441
show aspherical features on the images. Table~\ref{tab:objs} (Column~4)
lists the morphologies of all the control objects.

The DUSTY models used and the fitting procedures for the control objects 
were the same as those for the WFs. The only exceptions were
IRAS~07134$+$1005 and IRAS~22272$+$5435. 
Owing to their carbon-rich nature, the grain type of 90\% amorphous carbon
\citep{hanner88nasa} and 10\% SiC \citep{pegourie88aa} was used instead of
silicates for their DUSTY models. We have confirmed that as long as the optical
depth is not too large (e.g., $\tau_{2.2}<1$), the emission features 
of carbonaceous compounds will not
be prominent and hence the SED profile shape between the carbon- or oxygen-rich
envelopes are very similar.

\section{Results and Interpretations}
\label{sec:res}

\begin{table}
\caption{DUSTY parameters of each object which has 
a good model fit for the SED (See Table~\ref{tab:objs}).}
\label{tab:dusty}
\begin{tabular}{llrrr}
\hline


{Object} &
{Profile$^{\rm a}$} &
{$T_{\rm eff}$} &
{$T_{\rm d}$} &
{$\tau_{2.2}$} \\

&
&
{(K)} &
{(K)} &
\\

\hline
\multicolumn{5}{c}{Water Fountains} \\
\hline
IRAS~18113$-$2503 &  D? & 4400 & 100 & 0.700   \\
OH~12.8$-$0.9 &  S & 2200 & 700 & 3.000   \\
IRAS~18455$+$0448 &  S & 4400 & 600 & 2.000   \\
IRAS~19356$+$0754 &  D & 2000 & 200 & 2.000   \\
\hline
\multicolumn{5}{c}{Control AGB Stars} \\
\hline
IRAS~14247$+$0454 &  S & 2400 & 400 & 0.045   \\
IRAS~18556$+$0811 &  S & 2400 & 700 & 0.300   \\
IRAS~19149$+$1638 &  S & 2000 & 700 & 0.450   \\
IRAS~19312$+$1130 &  S & 2000 & 800 & 0.300   \\
IRAS~19395$+$1827 &  S & 2200 & 500 & 0.100   \\
IRAS~19495$+$0835 &  S & 2000 & 700 & 0.450   \\
\hline
\multicolumn{5}{c}{Control Post-AGB Stars} \\
\hline
IRAS~17534$+$2603 & S? & 3400 & 1600 & 1.000 \\
IRAS~20547$+$0247 & S? & 3600 & 900 & 4.000 \\
IRAS~22272$+$5435$^{\rm b}$ &  D & 8200 & 200 & 0.010   \\
\hline
\end{tabular}
\\
\flushleft

$^{\rm a}${SED profiles are classified into single-peaked ``S'' or
            double-peaked ``D''. An extra ``?'' means the SED is marginally
            fit by the model.}\\
$^{\rm b}${Carbon-rich object. The dust component parameters 
            used in the DUSTY model is different from other oxygen-rich 
            objects. See text for details.}\\

\end{table}

\begin{figure*}
   \includegraphics[scale=0.8]{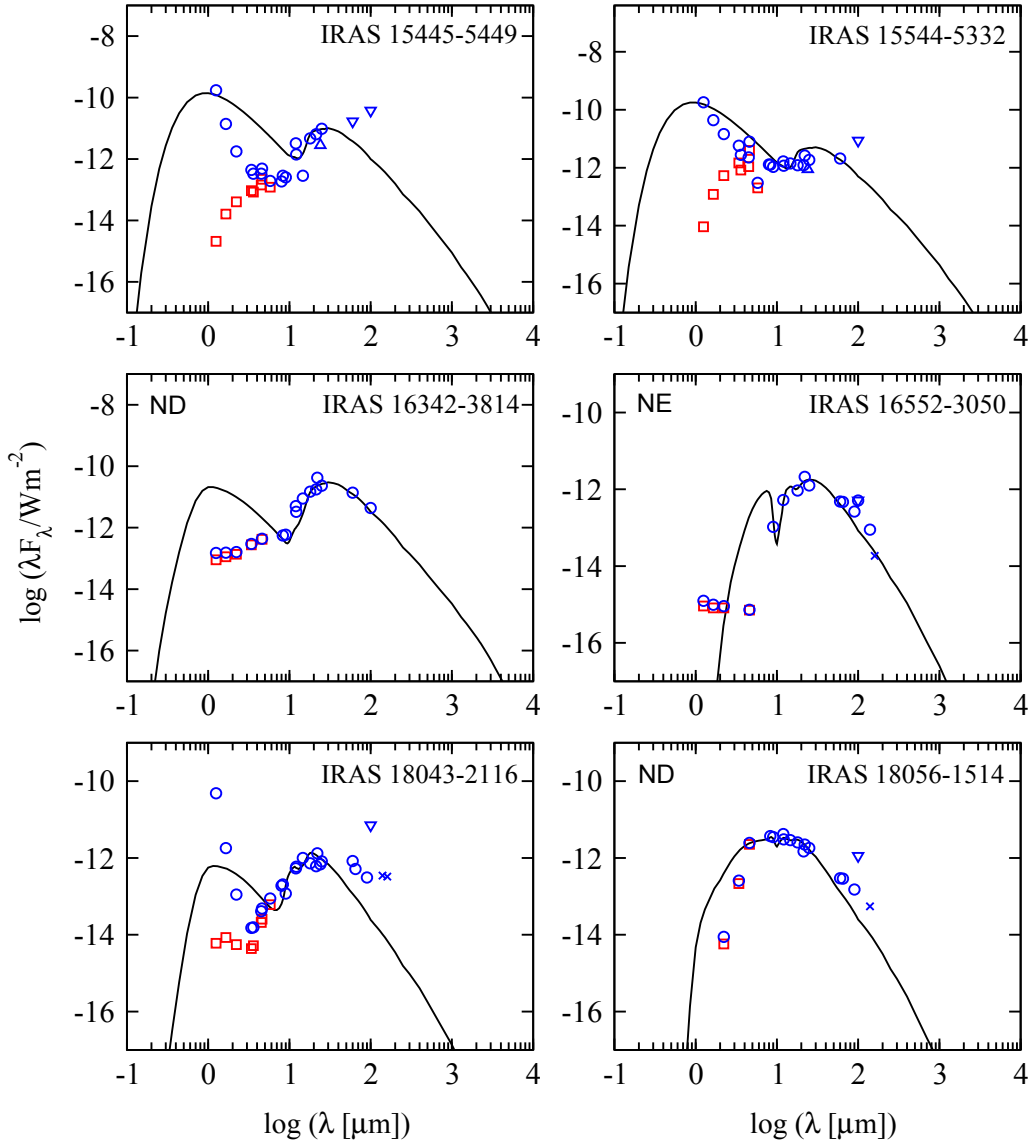}
   \caption{Spectral energy distributions of the water fountains.
             The notations used in this figure
             are the same as Figure~\ref{fig:sed_agb}.
            }
   \label{fig:sed_wf}
\end{figure*}

\begin{figure*}
   \includegraphics[scale=0.8]{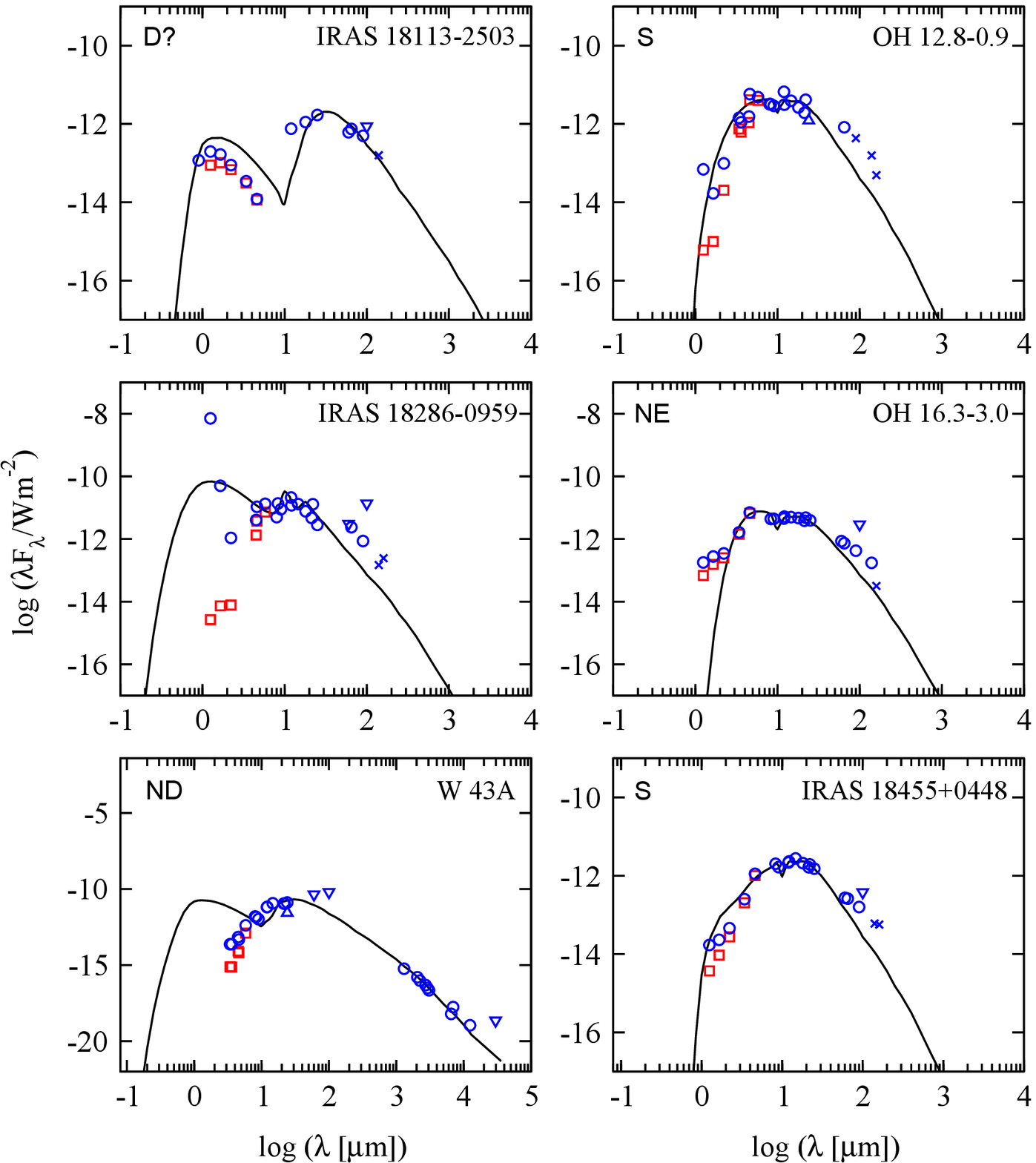}
   \contcaption{}
\end{figure*}

\begin{figure*}
   \includegraphics[scale=0.8]{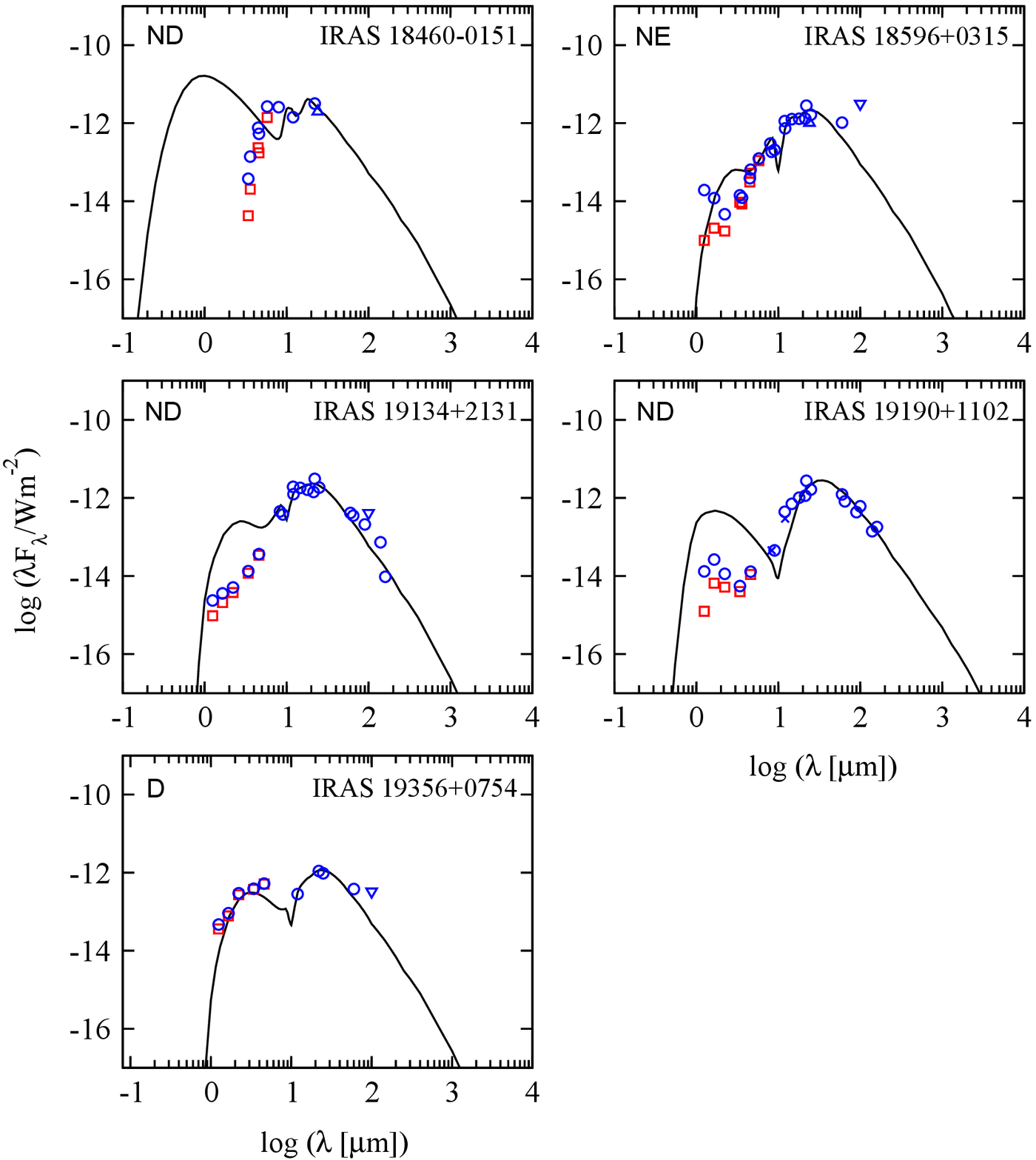}
   \contcaption{}
\end{figure*}


\subsection{Types of Profiles}
\label{ssec:typ}

There are mainly three types of SED profiles. 
\begin{enumerate}
\item[(1)] The envelope is spherical and the SED 
can be fit by a DUSTY model with either a single-peaked (denoted by ``S'' in
Tables~\ref{tab:objs} and \ref{tab:dusty}) or double-peaked (``D'') 
profile. A single-peaked profile means the envelope is not detached, e.g.,
in cases of usual AGB stars. A double-peaked profile means the 
envelope is detached, which is a characteristic of post-AGB stars. In such
a case the near-infrared peak is mainly 
contributed by the reddened photosphere of the central star, and the 
far-infrared peak is attributed to the cold dust in the detached envelope 
\citep{kwok93araa}.
\item[(2)] Some data points in mid- to far-infrared range can be fit by 
DUSTY, but the near-infrared data cannot be fit. Each of these 
SEDs either has near-infrared flux deficit (``ND'') or excess (``NE'') when 
comparing to the 
corresponding closest model curves. According to \citet{lagadec11mnras},
the near-infrared excess could be caused by the emission from hot dust
close to the star, and the envelope of such star is likely bipolar;
the near-infrared deficit could be a result of a thick torus that 
absorb radiation. In both cases the envelopes are aspherical and hence no
good fits can be obtained for the entire SEDs.
\item[(3)] This type of SEDs is similar to (2). The major difference 
is that in this case the near-infrared data points are not too
reliable due to heavy interstellar extinction, hence it is difficult to judge
whether they have infrared excess or deficit. These envelopes are also
likely aspherical. 
\end{enumerate}

\subsection{Control Objects}
\label{ssec:con}

All of our control AGB stars exhibit a single-peaked SED profile and good 
fits could be obtained from DUSTY models (Figure~\ref{fig:sed_agb}). 
Table~\ref{tab:dusty} 
gives the corresponding DUSTY parameters. 
The maximum flux is 
found in the near-infrared region, at about 2~$\mu$m. This is common for AGB 
stars according to previous studies \citep[e.g.,][]{groenewegen95aa}.
The temperatures for the best-fit
models are: 2000~K~$\leq$~$T_{\rm eff}$~$\leq$~2400~K;
400~K~$\leq$~$T_{\rm d}$~$\leq$~800~K. The optical depths
($\tau_{2.2}$) are mostly between 0.1--0.5. Usually the larger the optical 
depth is, the more prominent the 9~$\mu$m silicate emission feature is.

For the control post-AGB stars, 
there are no good fits for the SEDs of IRAS~07134$+$1005, OH~231.8$+$4.2, and
IRAS~17441$-$2411 (Figure~\ref{fig:sed_ppn}). This is not unexpected because 
their envelopes are aspherical as shown in the infrared image catalogue from
\citet{lagadec11mnras}. IRAS~17534$+$2603 and IRAS~20547$+$0247 are two
unresolved objects from the same catalogue which means they could have
spherical envelopes or they are too far away. Both SEDs could be marginally
fit by DUSTY with single-peaked models. It is possible that their 
envelopes have just started to depart from spherical symmetry, and hence
the aspherical features are not revealed in images.  
IRAS~22272$+$5435 is the only post-AGB star with a good fit here. The obtained
effective temperature ($T_{\rm eff}=8200$~K)
is higher than the AGB stars, but the dust temperatures ($T_{\rm d}=200$~K) 
is lower. 
This is expected for a post-AGB star where the central star is
becoming hotter, while the detached dust envelope is expanding and becoming
cooler. 
The obtained temperatures are similar to some post-AGB stars
modelled by DUSTY as presented in
\citet{suren02apss}. The best fit optical depth is 
$\tau_{2.2}=0.01$. Note that our $T_{\rm eff}$ is higher than the result
from \citet{szczerba97aa}, where they got $T_{\rm eff}=5300$~K. The
discrepancy is mainly due to the different extinction values used, which
affects the fluxes in short wavelengths that constraint the stellar 
temperature. \citet{szczerba97aa} adopted $A({\rm V})=1.0$ and $2.0$, while
we estimated $A({\rm V})=5.537$.
Since this object is known to have aspherical envelope
features revealed by CO (see Section~\ref{ssec:jus}), so it may seem odd that
a good fit is available, in particular when comparing with the similar case
IRAS~07134$+$1005. This can actually be explained 
by the fact that, via CO emission the existence of a spherical 
shell is predicted and the torus found in this object is several 
times smaller than that in IRAS~07134$+$1005 \citep{nakashima12apj}. Hence
it is likely that the molecular distribution in the outer envelope of 
IRAS~22272$+$5435 still remains mostly undisturbed, but this is not the case 
for IRAS~07134$+$1005. 

As a result, it can be seen that that our method is in general able to 
reproduce known results 
quite accurately: good fits for spherical envelopes, but not for aspherical 
envelopes. Nonetheless, we notice that the SEDs of envelopes with aspherical 
features could sometimes still be reproduced by one-dimensional models, 
depending on how large and significant the aspherical features are.
Therefore the logic is that, if a good fit is obtained then it is very likely 
that the envelope is spherical though it is not guaranteed; 
however, if no good fit could be obtained
then it is quite certain that the envelope is really aspherical, upon some
restrictions discussed in Section~\ref{ssec:rad}.

\subsection{Water Fountains}
\label{ssec:wf}

\subsubsection{Overview}
\label{sssec:ove}

The 17 WFs or WF candidates show a variety of SED profiles.
(Table~\ref{tab:objs}). 
Good DUSTY fits are obtained for four WF SEDs,
and they are classified 
into two categories: single-peaked or double-peaked profile.
Two of the SEDs could be fit by single-peaked
model curves (OH~12.8$-$0.9, and
IRAS~18455$+$0448). These two objects are found to be relatively young WF 
members, i.e., AGB or early post-AGB stars.
Another two WFs could arguably be fit by double-peaked curves 
(IRAS~18113$-$2503 and IRAS~19356$+$0754). These are likely to be post-AGB
stars. 
Some of the data points in these cases seem deviate from the model curves, 
but we notice that the curves still reasonably 
reproduce the SED line shapes, thus we treat them as cases that are fit. 

On the other hand,
as mentioned in Section~\ref{sec:intro}, a number of WFs are known to
have bipolar structures from infrared images. Those are the
cases that could not be fit by DUSTY, as expected.  
For these 13 SEDs without good fits, more than a few of them share a
similarity: there is a peak at mid- to far-infrared range that could be fit by 
DUSTY, but the near-infrared data cannot be fit. Each of these particular 
SEDs either has near-infrared deficit or excess when comparing to the 
corresponding closest model curves. All these cases without good fits are
likely to be post-AGB stars with aspherical envelope features. Note again
that there is strong near-infrared extinction toward some objects, and for 
those cases the fits are also not reliable.

Some objects have different flux values from similar wavelength
bands. A major reason is the difference in 
measured aperture size adopted from different instruments accounting for those
bands. Another
possible reason is that those objects may
have a certain degree of variability. Nonetheless, we found that the small
deviations do not affect our fitting results when there are enough data points
for constructing the SEDs.

Column 4 of 
Table~\ref{tab:objs} gives the morphologies of the objects according to
infrared images, if any; Column~5 gives a summary on whether the objects 
could be fit, and which categories that the objects have fallen into. 
Table~\ref{tab:dusty} lists the best-fit DUSTY parameters 
($T_{\rm eff}$, $T_{\rm d}$, and $\tau_{2.2}$) for the objects with 
good fits.
The SEDs and DUSTY model curves with the smallest sum of least-square
values are shown in Figure~\ref{fig:sed_wf}.

\subsubsection{Single-Peaked SED Profiles}
\label{sssec:single}

The SEDs of OH~12.8$-$0.9 and IRAS~18455$+$0448 could be
fit by single-peaked DUSTY models which look similar to the AGB 
stars, but they are still distinguishable 
because peaks of these WF SEDs are shifted to the mid-infrared region.
This shift is preliminary due to the large amount of cold dust in the thicker
envelopes, which also accounts for their large $J-K$ colour.
The fact that these two WFs could be fit by 
the one-dimensional code means that they are probably having envelopes which 
are close to spherical. Recall that WFs are objects associated with bipolar 
jets that can be traced by H$_{2}$O maser emission. Intuitively they must be
aspherical objects. However, the physical length of one side of the maser jets 
are typically of order $10^{2}$--$10^{3}$~AU as mentioned earlier, but
the envelope of a standard AGB star could have a radius of about
$10^{5}$~AU \citep{habing96aar}. Hence, if no other form of large aspherical 
feature is present (see Section~\ref{sec:dis}), it is possible that even 
though a WF jet has already formed, the outer part of the envelope, which 
can be observed in mid- to far-infrared, still remains spherical.

The relatively low $T_{\rm eff}$ (2200~K) of OH~12.8$-$0.9 
is similar to those of the control AGB stars, indicating 
this object is likely still in the AGB phase. The H$_{2}$O maser
velocity coverage of this object is relatively small within the 
WF class, $\sim$48\kms; and the three-dimensional jet velocity is estimated 
to be 58\kms\ \citep{boboltz07apj}, which is the slowest jet found in WFs. 
Jet acceleration was observed to be happening for this
object, and therefore it is suggested to be a relatively young WF
(under the assumption that jets may accelerate as they develop). 

IRAS~18455$+$0448 was reported as a low-velocity WF candidate with velocity
coverages $<$40\kms\ \citep{yung13apj}. Its 1612~MHz OH maser profile  
was analyzed by \citet{lewis01apj}, and found that the
double-peaked feature had been fading over a period of 10 years; this 
object was argued to be a very young post-AGB star. 
Its H$_{2}$O maser emission was 
further analysed by \citet{vlemmings14aa} via interferometric observations
which confirmed its WF status. 
IRAS~18455$+$0448 has a higher $T_{\rm eff}$ 
(4400~K) than 
the AGB stars, implying that it could be slightly more evolved,
which again agrees with the discussions in \citet{lewis01apj} and 
\citet{yung13apj}. Similar to OH~12.8$-$0.9, the H$_{2}$O maser kinematics 
and the SED profile of IRAS~18455$+$0448 give the same prediction that this 
object could be a younger member of the WF class.

\subsubsection{Doubled-Peaked SED Profiles}
\label{sssec:double}

There are two WFs with double-peaked SEDs that could be arguably 
fit by DUSTY. The double-peaked profile is a characteristic of post-AGB stars
as mentioned earlier.
Both WFs have larger optical depths than the
control AGB stars, which is another hint that they could be more
evolved objects with thicker envelopes.

IRAS~18113$-$2503 has the largest velocity coverage ($\sim$500\kms) of 
H$_{2}$O maser emission amongst all the WFs \citep{gomez11apj}. The 
three-dimensional jet velocity is not known, but from this large spectral
velocity coverage it is possible that this object has the fastest WF jet.
No OH maser observation has been done toward this object, so the kinematical
condition of the outer envelope is not known. We considered the SED of this 
object as marginally fit, because the model curve is able to reproduce the 
general line shape and the double-peaked feature with a brighter cold peak.
Nonetheless, the deviation from some data points indicates that the envelope 
is perhaps not totally spherical.
\citet{gomez11apj} proposed that this object is a post-AGB
star, based on the infrared characteristics.

IRAS~19356$+$0754 is a WF candidate that has a rather irregular OH maser 
profile \citep{yung14apj}. Its \textit{IRAS} colours suggest 
that it should be a (Mira) variable star at the end of the mass-loss phase 
which has a very thick oxygen-rich envelope 
\citep[Region~IV in][]{vanderveen88aa}. These properties show that the object 
should be a late/post-AGB star. However, the best fit DUSTY model for the 
SED gives $T_{\rm eff}\sim 2000$~K, which is unexpectedly low for such evolved 
object. It is possible that this star is undergoing a third dredge-up, which
would significantly cool down its effective temperature due to the sudden
increase of the atmospheric opacity \citep[e.g.,][]{herwig05araa}. More 
evidences are needed to confirm the true status of this object.

\subsubsection{SED Profiles without Good Fits}
\label{sssec:without}

No good fits from DUSTY models could be obtained for 13 
WFs or WF candidates. Some of them are known to have well-developed 
bipolar structures in infrared wavelengths, such as IRAS~15445$-$5449 
\citep{lagadec11mnras,sanchez11mnras,sanchez13mnras} and IRAS~16342$-$3814 
\citep{sahai99apj}. The deviation from the DUSTY best-fit model is mainly 
due to the non-spherical structure of the envelopes.
It is found that 
more than half of these unfit SEDs have near-infrared deficit when comparing 
to the model curves. 
The objects showing this feature are 
IRAS~15445$-$5449, IRAS~16342$-$3814, IRAS~18460$-$0151, 
IRAS~19314$+$2131, and IRAS~19190$+$1102 (see Figure~\ref{fig:sed_wf}). 
Using IRAS~16342$-$3814 as a case study, 
\citet{murakawa12aa} suggested that this type of SEDs could be reproduced by
a model consists of a spherical AGB envelope, an optically thick torus, and
a bipolar jet. This is a typical post-AGB star structure, and that means most
of these unfit WFs could have already evolved into the post-AGB phase, 
except for the known example of W~43A \citep{imai02nature} and a possibly new
example IRAS~18056$-$1514 (see below), which seem to be AGB stars with
aspherical envelops.

The three objects IRAS~16552$-$3050, OH~16.3$-$3.0, and IRAS~18596$+$0315 
are a bit different 
from the above. The near-infrared fluxes of the former two objects reveal 
plateau features which are brighter than the model curves 
(see Figure~\ref{fig:sed_wf}). This near-infrared excess could be a result of
hot dust, the existence of 
circumstellar disks \citep[e.g.,][]{gezer15mnras}, or from possible
hot companions. 
IRAS~18596$+$0315 shows not a plateau in near-infrared range, but an increasing 
slope toward the shorter wavelength (Figure~\ref{fig:sed_wf}). It is unsure 
whether this feature is reliable due to strong extinction. Its WF status was
first suggested by \citep{engels02aa}. An interferometric observation of
the OH maser reveals a slightly bipolar structure \citep{amiri11aa}, and again
this is a signature of the envelope departing from spherical symmetry.

It is difficult to judge whether IRAS~15544$-$5332 has a good SED fit or not.
The reason is that most of the reliable data points are clustered in a 
relatively
small mid-infrared range, and the near-infrared data points are not quite
reliable due to strong extinction (Figure~\ref{fig:sed_wf}). Therefore the
result is inconclusive but we conservatively put it as an unfit case. 
This is a WF candidate with only one H$_{2}$O maser emission peak found 
outside the velocity coverage of its OH maser, which has a slightly irregular
double-peaked profile \citep{deacon07apj}. An irregular OH maser profile 
indicates that the
outer envelope has been disturbed (e.g., by jets) so that it is no longer 
spherical \citep{zijlstra01mnras}. In this case, the envelope might have just 
started to depart from spherical symmetry, thus the double-peaked OH maser 
profile has not been totally broken down. The SED profile tells a similar
story, as the merely-fit model curve could imply that the 
possibly detached/detaching AGB envelope is about to become aspherical.

IRAS~18056$-$1514 is an odd case here. No good fit is obtained but the
closest model curve is single-peaked and looks like that for 
OH~12.8$-$0.9.
There is no further study
regarding IRAS~18056$-$1514, and in fact its WF status remains very doubtful. 
It is because its only H$_{2}$O maser emission peak that was found outside the
OH velocity range \citep{yung13apj}, was not present in a later observation 
\citep{yung14apj}. This object is likely to be 
a late AGB star according to the infrared colours, and this agrees with its
single-peaked SED profile.

\section{Discussion}
\label{sec:dis}

We have shown that good DUSTY fits could be obtained for some WF SEDs which
imply spherical envelopes, but others are likely associated with aspherical
envelopes. The results suggest that
even though all WFs possess maser jets with apparently similar
dynamical ages, the objects could be coming from different stages of 
morphological metamorphosis. Since this change in morphology is known to be
closely related to the evolutionary status, our results also 
imply the fact that WFs could have different evolutionary statuses. This
fact agrees with other evidences such as the colours of the objects.
Note that it might be difficult to determine whether the SED fits are good for 
several WFs, but the remaining unambiguous cases are sufficient to support
the above ideas. 
There are several details require deeper exploration regarding the role of
WFs. In this
section we will discuss the problem about the definition of dynamical ages,
and also the possible relationship between WF jets and other types of 
outflows.

\subsection{Dynamical Age}
\label{ssec:age}

The main reason that WFs are being regarded as objects started to deviate 
from spherical symmetry is the short dynamical
ages of their jets. This contradicts to the results of the present study. 
If WF jets really indicate the earliest moment of such
envelope morphological change, we expect that a majority of WFs would still
have rather spherical envelope. It is because the WF jets are small in scale
(see Section~\ref{sec:intro}) and hence they probably could not disturb the
outer infrared envelope so much within less than 100 years.  
We suggest that this contradiction arisen from the method that has been used 
to define and estimate the dynamical ages of WF jets. 

The usual method is to perform multi-epoch
very long baseline interferometry observations to measure the proper motions
on the sky plane of individual H$_{2}$O maser features 
\citep[see, e.g.,][]{imai07apj,yung11apj}.  
Then the jet dynamical age is 
estimated from the observed spatial extent of the jet (traced by the maser
feature distribution) and the jet velocity on the sky plane. However, whether 
this is the real
age of the jet is not guarantee. This method of estimation
assumes that the maser features are able to trace the tip of the jet.
In fact, the maser features could be tracing only the innermost part of
the entire jet. It is possible that the tip of the jet have 
arrived at the more outer region of the envelope, but cannot be seen in
maser observations. This could happen because H$_{2}$O molecules are more 
abundant in the inner region (within hundreds of AU from the central star) 
of the envelope and hence easier to be observed via maser emission.

The dynamical age calculated with the above method therefore does not have a 
real physical meaning, and hence not all WFs are 
really the ``transitional objects'' that we have been looking for. 
Furthermore, it is actually questionable whether 
the larger jets or bipolar structures observed in infrared images 
\citep[e.g.,][]{lagadec11mnras} must be developed from smaller jets such as the 
WF jets. Nonetheless, this class of object is still very 
valuable because the maser emission allows us to inspect deep into the
root of the bipolar jets, which could provide essential physical constraints
for understanding the jet formation mechanism.

\subsection{Torus}
\label{ssec:tor}


\begin{figure}
   \includegraphics[scale=0.9]{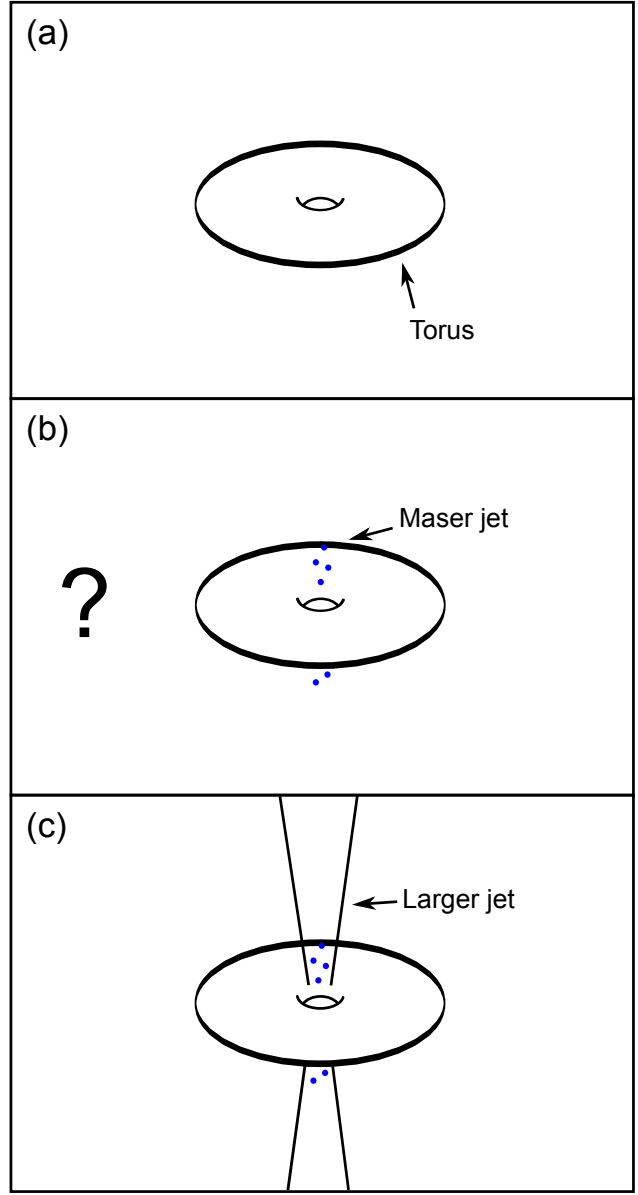}
   \caption{Schematic view of a possible evolutionary sequence starts with 
             (a) the formation of a torus, followed by (b) a water fountain 
             maser jet, and then (c) a larger jet. Note that it is
             unclear whether the small-scale maser jet must exist earlier
             than the larger jet, or they could coexist from the beginning.
             Step (b) might be omitted. 
             }
   \label{fig:schematic}
\end{figure}


According to our results, WFs could be objects from different morphological
metamorphosis stages. In this subsection we try to speculate how this could
be possible.
We should note that jets are not the only type of aspherical outflow that 
affects an envelope's morphology.
Observations have shown that an AGB/post-AGB star with bipolar jet is very 
likely to be associated with a torus. The torus is suggested to be appeared
quite ``suddenly'' within a short time, and the jet is formed 
almost simultaneously or shortly after the formation of the torus 
\citep[e.g., by a few hundred years,][]{huggins07apj}. 
Here the ``jet'' does not refer only to 
the specific small-scale WF jets, but also stellar jets in general which could 
be much larger in scale. Such a torus-jet configuration is currently best
described by the existence of a binary component around the primary 
AGB/post-AGB star \citep[see,][and the references therein]{huggins07apj}.

If this is the case, then the WFs are expected to be associated with torii
as well. In fact, the existence of torii has been proposed for 
W~43A \citep{imai05apj} according to the distribution of SiO maser, and for 
IRAS~16342$-$3814 \citep{verhoelst09aa} because of the 
``dark waist'' observed in mid-infrared images. IRAS~18286$-$0959 has a few
outlier H$_{2}$O maser features observed in the equatorial region
\citep{imai13apj1}, which might be related to a torus as well.
There is no such evidence found for other WFs so far, however, we suggested
that some of the unfit SED could be resulted from a well-developed torus
feature. As discussed in Section~\ref{ssec:con}, both of our control post-AGB
stars IRAS~07134$+$1005 and IRAS~22272$+$5435 are known to have a torus but
only the former has an unfit SED by DUSTY. This is very likely related to the
size of their corresponding torii. In addition, the near-infrared plateau
features observed in some WFs mentioned in Section~\ref{sssec:without} could 
also be related to disk-like features \citep{gezer15mnras}, and/or hot
companions.

Figure~\ref{fig:schematic} shows a possible scenario putting the torus,
WF maser jet, and the larger jet together. Note that even though it has been
vastly believed that jets grow from smallest sizes 
(see Section~\ref{sec:intro}), there is in fact no concrete evidence 
showing that this is the only possible way for jets to develop. The only
evidence is perhaps the existence of WFs with spherical envelops 
(Section~\ref{sssec:single}) which show that 
no larger jet structure is present yet, but this evidence alone is not strong
enough to be conclusive. 
Therefore,
step (b) shown in Figure~\ref{fig:schematic} may not be necessary. If this
step does exist, then some of the WFs may really be objects possessing young
jets; if this step does not exist, then there is no way to tell from the
WF jets the real age of the entire jet. In either case, the WF jets trace
the innermost part of the entire jet in this scenario.

\subsection{Extreme-Outflow}
\label{ssec:ext}

Another type of outflow is the molecular jets that can usually be
traced by thermal line emission from CO molecules, 
to which \citet{sahai15apjl} has coined the term ``extreme-outflow''. These 
jets could have velocities $>$100\kms, similar to the WF jets, but with 
larger physical sizes (see Section~\ref{sec:intro}). The relationship
between WF jets and extreme-outflows is unclear because so far the latter
has not been detected toward the WFs, and this is probably due to sensitivity 
constraints and foreground/background contaminations \citep{rizzo13aa}. 
However, a comparison on the jet orientations of WF jets, extreme-outflows and 
the large infrared bipolar structures may give us some hints. 

It is found that the orientations of the extreme-outflows align with the 
possibly bipolar envelopes 
of the corresponding objects as shown in the optical/infrared images 
\citep{olofsson15aa,sahai15apjl}.
The estimated momentum of the outflow was too large to be supported by
radiation pressure from the central star, and hence a mechanism driven by
binary system is more plausible \citep{sahai15apjl}, similar to that
discussed in Section~\ref{ssec:tor}.
Therefore it is likely that the extreme-outflow has direct connection to the
shaping of such aspherical envelopes. Similarly, for the WFs
IRAS~15445$-$5449 \citep{sanchez13mnras}, IRAS~16342$-$3814 
\citep{sahai99apj}, and W~43A \citep{imai02nature}, the spatial extent of
their envelopes in infrared images (even in optical for the case of 
IRAS~16342$-$3814) align very well with their corresponding WF maser jets.
Other WFs are either unobserved or unresolved in infrared imaging. 

From these comparisons, it is not surprised that both extreme-outflows and
WF jets have a connection to the large bipolar structures. An intuitive
guess is that these two types of jets could even be aligned. However, 
we arrive at the same problem, i.e., whether the smaller WF jets are the 
younger version of the larger extreme-outflows. If this is true, then the
extreme-outflow might represent a chronological stage somewhere between
(b) and (c) in Figure~\ref{fig:schematic}, but we have no sufficient
information to examine this.

\section{Conclusions}
\label{sec:con}

The various morphologies of PNe are suggested to be shaped by high velocity
bipolar jets from (late) AGB and post-AGB stars. 
It has been widely believed that such jets are developed from tiny young jets
such as those from the WFs. Hence WFs are regarded as objects representing
the onset of the morphological change of envelopes. However, there is no
concrete evidence supporting this hypothesis.
We performed a collective study of SED profiles by fitting one-dimensional 
dust radiative transfer models generated by the DUSTY code. Our objects 
included the confirmed WFs, as well as a few WF candidates 
(17 objects in total). We have 
also studied some known AGB and post-AGB stars as control objects.
Our findings are summarized as follow.
\begin{enumerate}
\item The SED profiles of two WFs could be fit by single-peaked DUSTY 
      models which are all peaked in the mid-infrared range. Another two could
      arguably be  
      fit by double-peaked models resembling those for the post-AGB stars. 
      No good fits could be obtained for the remaining 13 WFs; more than half 
      of these objects exhibit double-peaked line shapes with
      the near-infrared peak significantly weaker than the far-infrared peak.
      The results confirm that WFs could possess different envelope 
      morphologies and hence it is unlikely that they must be objects just
      started to deviate from spherical symmetry as previous believed. 
\item The short dynamical age of WFs have no real physical meaning, and it
      cannot be used as an evidence to claim that WF jets must be young. WF
      jets could be the innermost part of well-developed larger jets.
\item The role of the WF jets are discussed together with other common 
      aspherical envelope components such as torii and the molecular jets
      revealed by CO thermal line emission (the ``extreme-outflows''). 
      According to existing observations and theories, it is
      likely that the torii will be formed ahead of the jets, but whether 
      there is any chronological relationship between the WF jets and the 
      extreme-outflow is unclear. Both types of jets are shown to be
      correlated with the large-scale bipolar structure revealed in infrared
      observations for some objects. 
\end{enumerate}
The present work is also the pioneer collective study of WF envelopes
by simple but effective one-dimensional radiative transfer models. We suggest 
that this approach will be useful 
for future statistical studies of larger number of stellar maser sources.

\section*{Acknowledgements}

This work is supported by Act 211 Government of the Russian Federation, 
agreement No. 02.A03.21.0006a. J.N. is supported by grants from the Research
Grants Council of Hong Kong (project code: HKU 704411P), and the Small
Project Funding of the University of Hong Kong (project code:
201007176004). CHH is supported by the Small Project Funding of the 
University of Hong Kong (project No. 201007176028), and the Science and 
Technology Development Fund of Macau 
(project code: 039/2013/A2 and 017/2014/A1).
The Laboratory for Space Research was established by a special grant from 
the University Development Fund of The University of Hong Kong. This work 
is also in part supported by grants 
from the HKRGC (HKU 7027/11P and HKU 7062/13P).
This publication makes use of data products from the 
\textit{Two Micron All Sky Survey}, which is a joint project of the University 
of Massachusetts and the Infrared Processing and Analysis Center/California 
Institute of Technology, funded by the National Aeronautics and Space 
Administration and the National Science Foundation; 
\textit{Wide-field Infrared Survey Explorer}, which is a joint project of the 
University of California, Los Angeles, and the Jet Propulsion 
Laboratory/California Institute of Technology, funded by the National 
Aeronautics and Space Administration;
\textit{Midcourse Space Experiment}, processing of the data was funded by the 
Ballistic Missile Defense Organization with additional support from NASA 
Office of Space Science; NASA/IPAC 
Infrared Science Archive, which is operated by the Jet Propulsion Laboratory, 
California Institute of Technology, under contract with the National 
Aeronautics and Space Administration;
\textit{Infrared Astronomical Satellite} ~(\textit{IRAS}), a joint project 
of the US, UK and the Netherlands;
\textit{AKARI}, a JAXA project with the participation of ESA;
\textit{Spitzer Space Telescope}, which is operated by the Jet Propulsion 
Laboratory, California Institute of Technology under a contract with NASA.
The National Radio Astronomy Observatory is a facility of the National Science
Foundation operated under cooperative agreement by Associated
Universities, Inc.
The Nobeyama Millimeter Array was operated by Nobeyama Radio Observatory, 
a branch of National Astronomical Observatory of Japan, National Institutes 
of Natural Sciences.






\bibliographystyle{mnras}
\bibliography{ms} 




\appendix

\section{Photometric Data}
\label{app:data}

The original sets of photometric data used in the current analysis are given
in Tables~\ref{tab:2mass}--\ref{tab:spitzer}.

\begin{table*}
\caption{\textsl{2MASS} photometric data and the $A({\rm V})$ colour excess 
of each object (see text).}
\label{tab:2mass}
\flushleft
\begin{tabular}{lrrrrrrr}
\hline
{Object} &
{$A({\rm V})$} &
{J-band} &
{$\sigma_{\rm J}$} &
{H-band} &
{$\sigma_{\rm H}$} &
{K-band} &
{$\sigma_{\rm K}$} \\

&
&
{(mag)} &
{(mag)} &
{(mag)} &
{(mag)} &
{(mag)} &
{(mag)} \\
\hline

\multicolumn{8}{c}{Water Fountains} \\
\hline
IRAS~15445$-$5449 & 0.056 & 15.693 & 0.080 & 12.701 & 0.040 & 10.889 & 0.024 \\
IRAS~15544$-$5332 & 44.166 & 14.265 & 0.063 & 12.465 & $\cdots$ & 11.355 & $\cdots$ \\
IRAS~16342$-$3814 & 1.996& 11.608 & 0.033 & 10.589 & 0.038 & 9.569 & 0.027 \\
IRAS~16552$-$3050 & 1.231 & 16.594 & 0.181 & 15.945 & 0.209 & 15.125 & $\cdots$ \\
IRAS~18043$-$2116 & 35.011 & 14.546 & 0.049 & 13.404 & 0.062 & 13.042 & 0.065 \\
IRAS~18056$-$1514 & 4.951 & 11.334 & 0.023 & 10.028 & 0.023 & 9.509 & 0.021 \\
IRAS~18113$-$2503 & 3.143 & 11.630 & 0.033 & 10.698 & 0.034 & 10.323 & 0.030 \\
OH~12.8$-$0.9 & 18.461 & 17.041 & $\cdots$ & 15.725 & $\cdots$ & 11.639 & 0.025 \\
IRAS~18286$-$0959 & 57.614 & 15.431 & 0.097 & 13.562 & 0.065 & 12.674 & 0.060 \\
OH~16.3$-$3.0 & 3.748 & 11.905 & 0.024 & 10.247 & 0.023 & 8.904 & 0.021 \\
W~43A & 110.537 & 16.361 & 0.126 & 14.988 & 0.079 & 14.120 & 0.088 \\
IRAS~18455$+$0448 & 5.918 & 15.068 & 0.056 & 13.294 & 0.030 & 11.305 & 0.023 \\
IRAS~18460$-$0151 & 61.544 & 13.732 & $\cdots$ & 13.813 & 0.057 & 13.435 & 0.055 \\
IRAS~18596$+$0315 & 11.560 & 16.501 & $\cdots$ & 14.945 & 0.088 & 14.311 & $\cdots$ \\
IRAS~19134$+$2131 & 3.525 & 16.543 & 0.128 & 14.926 & 0.071 & 13.464 & 0.038 \\
IRAS~19190$+$1102 & 9.142 & 16.253 & 0.118 & 13.690 & $\cdots$ & 13.120 & $\cdots$ \\
IRAS~19356$+$0754 & 1.017 & 11.376 & 0.024 & 10.770 & 0.024 & 10.555 & 0.021 \\
\hline
\multicolumn{8}{c}{AGB Stars} \\
\hline
IRAS~14247$+$0454 & 0.090 & 3.091 & 0.246 & 2.229 & 0.266 & 1.532 & 0.298 \\
IRAS~18556$+$0811 & 7.961 & 7.162 & 0.024 & 5.264 & 0.026 & 3.953 & 0.036 \\
IRAS~19149$+$1638 & 8.038 & 10.107 & 0.023 & 7.564 & 0.031 & 5.889 & 0.027 \\
IRAS~19312$+$1130 & 1.442 & 9.044 & 0.022 & 7.292 & 0.055 & 6.028 & 0.018 \\
IRAS~19395$+$1827 & 4.030 & 8.049 & 0.018 & 6.314 & 0.026 & 5.243 & 0.017 \\
IRAS~19495$+$0835 & 0.533 & 6.487 & 0.026 & 4.769 & 0.034 & 3.696 & 0.274 \\
\hline
\multicolumn{8}{c}{Post-AGB Stars} \\
\hline
IRAS~07134$+$1005 & 0.248 & 6.868 & 0.021 & 6.708 & 0.036 & 6.606 & 0.017 \\
IRAS~19024$+$0044 & 4.104 & 12.404 & 0.026 & 11.519 & 0.024 & 10.763 & 0.023 \\
IRAS~22272$+$5435 & 5.537 & 5.371 & 0.020 & 4.894 & 0.029 & 4.508 & 0.016 \\
\hline
\end{tabular}
\end{table*}


\begin{table*}
\caption{\textsl{WISE} photometric data.}
\label{tab:wise}
\flushleft
\begin{tabular}{lrrrrrrrr}
\hline
{Object} &
{$3.4~\mu$m} &
{$\sigma_{3.4\mu{\rm m}}$} &
{$4.6~\mu$m} &
{$\sigma_{4.6\mu{\rm m}}$} &
{$12~\mu$m} &
{$\sigma_{12\mu{\rm m}}$} &
{$22~\mu$m} &
{$\sigma_{22\mu{\rm m}}$} \\

&
{(mag)} &
{(mag)} &
{(mag)} &
{(mag)} &
{(mag)} &
{(mag)} &
{(mag)} &
{(mag)} \\

\hline

\multicolumn{9}{c}{Water Fountains} \\
\hline
IRAS~15445$-$5449 & 8.883 & 0.023 & 7.049 & 0.019 & 1.607 & 0.013 & $\cdots$ & $\cdots$ \\
IRAS~15544$-$5332 & 5.889 & 0.040 & 3.907 & 0.054 & 2.326 & 0.015 & $-$0.308 & 0.016 \\
IRAS~16342$-$3814 & 7.701 & 0.017 & 6.343 & 0.015 & 0.266 & 0.006 & $-$3.323 & 0.001 \\
IRAS~16552$-$3050 & $\cdots$ & $\cdots$ & 15.021 & 0.292 & 6.697 & 0.019 & 5.631 & $\cdots$ \\
IRAS~18043$-$2116 & 12.203 & 0.100 & 9.382 & 0.028 & 3.545 & 0.015 & 0.443 & 0.009 \\
IRAS~18056$-$1514 & 7.746 & 0.023 & 4.233 & 0.041 & 0.599 & 0.011 & $-$0.746 & 0.009 \\
IRAS~18113$-$2503 & 10.078 & 0.025 & 10.249 & 0.028 & $\cdots$ & $\cdots$ & $\cdots$ & $\cdots$ \\
OH~12.8$-$0.9 & 6.623 & 0.032 & 3.854 & 0.054 & 0.784 & 0.020 & $-$0.806 & 0.014 \\
IRAS~18286$-$0959 & $\cdots$ & $\cdots$ & 3.962 & 0.044 & $-$0.441 & 0.062 & $-$2.036 & 0.008 \\
OH~16.3$-$3.0 & 5.631 & 0.063 & 2.991 & 0.072 & 0.160 & 0.023 & $-$1.601 & 0.009 \\
W~43A & 13.879 & 0.347 & 10.597 & 0.051 & 0.451 & 0.413 & $-$2.574 & 0.014 \\
IRAS~18455$+$0448 & 7.780 & 0.025 & 5.080 & 0.031 & 1.306 & 0.010 & $-$0.607 & 0.006 \\
IRAS~18460$-$0151 & 11.610 & 0.052 & 7.017 & 0.020 & 0.824 & 0.021 & $-$1.069 & 0.010 \\
IRAS~18596$+$0315 & 11.367 & 0.030 & 8.613 & 0.022 & 2.696 & 0.014 & $-$0.384 & 0.015 \\
IRAS~19134$+$2131 & 11.128 & 0.023 & 9.068 & 0.020 & 2.107 & 0.009 & $-$0.480 & 0.011 \\
IRAS~19190$+$1102 & 12.300 & 0.261 & 10.301 & 0.032 & 3.717 & 0.015 & $-$0.351 & 0.014 \\
IRAS~19356$+$0754 & 6.429 & 0.046 & 5.137 & 0.034 & 3.349 & 0.015 & $-$0.058 & 0.018 \\
\hline
\multicolumn{9}{c}{AGB Stars} \\
\hline
IRAS~14247$+$0454 & 1.287 & $\cdots$ & 1.300 & 0.179 & $-$0.342 & 0.034 & $-$2.024 & 0.001 \\
IRAS~18556$+$0811 & 2.683 & 0.011 & 1.675 & 0.378 & $-$1.102 & 0.251 & $-$2.303 & 0.002 \\
IRAS~19149$+$1638 & 3.521 & 0.130 & 1.999 & 0.010 & 0.450 & 0.013 & $-$0.722 & 0.009 \\
IRAS~19312$+$1130 & 4.098 & 0.094 & 2.683 & 0.087 & 1.135 & 0.007 & 0.099 & 0.015 \\
IRAS~19395$+$1827 & 5.034 & 0.064 & 3.996 & 0.022 & 1.971 & 0.012 & 0.639 & 0.017 \\
IRAS~19495$+$0835 & 3.120 & 0.119 & 1.789 & 0.012 & $-$0.529 & 0.016 & $-$1.731 & 0.006 \\
\hline
\multicolumn{9}{c}{Post-AGB Stars} \\
\hline
IRAS~07134$+$1005 & 6.335 & 0.042 & 6.198 & 0.022 & $-$0.278 & 0.009 & $-$2.805 & 0.002 \\
IRAS~19024$+$0044 & 9.945 & 0.024 & 8.673 & 0.020 & 2.325 & 0.007 & $-$1.655 & 0.010 \\
IRAS~22272$+$5435 & 4.845 & 0.090 & 3.559 & 0.048 & $-$1.054 & 0.175 & $-$3.892 & 0.000 \\
\hline
\end{tabular}
\end{table*}


\begin{table*}
\caption{\textsl{IRAS} photometric data.}
\label{tab:iras}
\flushleft
\begin{tabular}{lrrrrrrrr}
\hline
{Object} &
{$12~\mu$m} &
{$\sigma_{12\mu{\rm m}}$} &
{$25~\mu$m} &
{$\sigma_{25\mu{\rm m}}$} &
{$60~\mu$m} &
{$\sigma_{60\mu{\rm m}}$} &
{$100~\mu$m} &
{$\sigma_{100\mu{\rm m}}$} \\

&
{(Jy)} &
{(\%)} &
{(Jy)} &
{(\%)} &
{(Jy)} &
{(\%)} &
{(Jy)} &
{(\%)} \\

\hline

\multicolumn{9}{c}{Water Fountains} \\
\hline
IRAS~15445$-$5449 & 6.88$\phantom{^{\dagger}}$ & 9 & 87.20 & 10 & 1130.00$^{\dagger}$ & $\cdots$ & 2180.00$^{\dagger}$ & $\cdots$ \\
IRAS~15544$-$5332 & 4.64$\phantom{^{\dagger}}$ & 13 & 15.50 & 12 & 41.50$\phantom{^{\dagger}}$ & 17 & 288.00$^{\dagger}$ & $\cdots$ \\
IRAS~16342$-$3814 & 16.20$\phantom{^{\dagger}}$ & 3 & 200.00 & 3 & 290.00$\phantom{^{\dagger}}$ & 8 & 139.00$\phantom{^{\dagger}}$ & 8 \\
IRAS~16552$-$3050 & 2.46$\phantom{^{\dagger}}$ & 5 & 10.50 & 8 & 9.58$\phantom{^{\dagger}}$ & 10 & 17.00$^{\dagger}$ & $\cdots$ \\
IRAS~18043$-$2116 & 6.60$^{\dagger}$ & $\cdots$ & 6.76 & 10 & 16.60$\phantom{^{\dagger}}$ & 16 & 237.00$^{\dagger}$ & $\cdots$ \\
IRAS~18056$-$1514 & 13.60$\phantom{^{\dagger}}$ & 16 & 15.20 & 17 & 5.94$\phantom{^{\dagger}}$ & 12 & 37.80$^{\dagger}$ & $\cdots$ \\
IRAS~18113$-$2503 & 2.90$\phantom{^{\dagger}}$ & 10 & 14.80 & 8 & 12.90$\phantom{^{\dagger}}$ & 17 & 29.20$^{\dagger}$ & $\cdots$ \\
OH~12.8$-$0.9 & 11.60$\phantom{^{\dagger}}$ & 6 & 16.90 & 8 & 13.90$\phantom{^{\dagger}}$ & 16 & 289.00$^{\dagger}$ & $\cdots$ \\
IRAS~18286$-$0959 & 24.90$\phantom{^{\dagger}}$ & 5 & 24.50 & 8 & 18.40$^{\dagger}$ & $\cdots$ & 405.00$^{\dagger}$ & $\cdots$ \\
OH~16.3$-$3.0 & 18.10$\phantom{^{\dagger}}$ & 8 & 30.10 & 13 & 16.90$\phantom{^{\dagger}}$ & 18 & 131.00$^{\dagger}$ & $\cdots$ \\
W~43A & 23.70$\phantom{^{\dagger}}$ & 4 & 104.00 & 5 & 295.00$^{\dagger}$ & $\cdots$ & 2520.00$^{\dagger}$ & $\cdots$ \\
IRAS~18455$+$0448 & 9.35$\phantom{^{\dagger}}$ & 6 & 12.60 & 6 & 5.47$\phantom{^{\dagger}}$ & 9 & 12.60$^{\dagger}$ & $\cdots$ \\
IRAS~18460$-$0151 & 20.90$^{\dagger}$ & $\cdots$ & 32.70 & 8 & 277.00$^{\dagger}$ & $\cdots$ & 291.00$\phantom{^{\dagger}}$ & 17 \\
IRAS~18596$+$0315 & 2.60$\phantom{^{\dagger}}$ & 9 & 14.20 & 8 & 22.60$\phantom{^{\dagger}}$ & 11 & 113.00$^{\dagger}$ & $\cdots$ \\
IRAS~19134$+$2131 & 5.06$\phantom{^{\dagger}}$ & 4 & 15.60 & 5 & 8.56$\phantom{^{\dagger}}$ & 9 & 3.95$^{\dagger}$ & $\cdots$ \\
IRAS~19190$+$1102 & 1.59$\phantom{^{\dagger}}$ & 7 & 13.70 & 6 & 24.50$\phantom{^{\dagger}}$ & 16 & 20.40$\phantom{^{\dagger}}$ & 15 \\
IRAS~19356$+$0754 & 1.12$\phantom{^{\dagger}}$ & 7 & 7.99 & 7 & 7.62$\phantom{^{\dagger}}$ & 10 & 10.90$^{\dagger}$ & $\cdots$ \\
\hline
\multicolumn{9}{c}{AGB Stars} \\
\hline
IRAS~14247$+$0454 & 109.00$\phantom{^{\dagger}}$ & 10 & 65.20 & 7 & 11.90$\phantom{^{\dagger}}$ & 9 & 4.29$\phantom{^{\dagger}}$ & 12 \\
IRAS~18556$+$0811 & 104.00$\phantom{^{\dagger}}$ & 7 & 84.70 & 7 & 10.20$\phantom{^{\dagger}}$ & 8 & 79.90$^{\dagger}$ & $\cdots$ \\
IRAS~19149$+$1638 & 14.50$\phantom{^{\dagger}}$ & 4 & 14.80 & 5 & 3.10$\phantom{^{\dagger}}$ & 21 & 13.20$^{\dagger}$ & $\cdots$ \\
IRAS~19312$+$1130 & 9.95$\phantom{^{\dagger}}$ & 12 & 7.63 & 7 & 1.29$\phantom{^{\dagger}}$ & 8 & 2.68$^{\dagger}$ & $\cdots$ \\
IRAS~19395$+$1827 & 8.35$\phantom{^{\dagger}}$ & 6 & 6.73 & 7 & 1.53$\phantom{^{\dagger}}$ & 16 & 21.60$^{\dagger}$ & $\cdots$ \\
IRAS~19495$+$0835 & 80.00$\phantom{^{\dagger}}$ & 5 & 65.50 & 10 & 11.80$\phantom{^{\dagger}}$ & 17 & 3.73$\phantom{^{\dagger}}$ & 13 \\
\hline
\multicolumn{9}{c}{Post-AGB Stars} \\
\hline
IRAS~07134$+$1005 & 24.50$\phantom{^{\dagger}}$ & 5 & 117.00 & 4 & 50.10$\phantom{^{\dagger}}$ & 26 & 18.70$\phantom{^{\dagger}}$ & 10 \\
IRAS~19024$+$0044 & 2.86$\phantom{^{\dagger}}$ & 6 & 48.80 & 8 & 42.50$\phantom{^{\dagger}}$ & 13 & 15.70$^{\dagger}$ & $\cdots$ \\
IRAS~22272$+$5435 & 73.90$\phantom{^{\dagger}}$ & 3 & 302.00 & 3 & 96.60$\phantom{^{\dagger}}$ & 10 & 41.00$^{\dagger}$ & $\cdots$ \\
\hline
\end{tabular}
\\
\flushleft
$^{\dagger}${Upper band flux limit.}
\end{table*}


\begin{table*}
\caption{\textsl{MSX} photometric data.}
\label{tab:msx}
\flushleft
\begin{tabular}{lrrrrrrrr}
\hline
{Object} &
{$8.28~\mu$m} &
{$\sigma_{8.28\mu{\rm m}}$} &
{$12.13~\mu$m} &
{$\sigma_{12.13\mu{\rm m}}$} &
{$14.65~\mu$m} &
{$\sigma_{14.65\mu{\rm m}}$} &
{$21.34~\mu$m} &
{$\sigma_{21.34\mu{\rm m}}$} \\

&
{(Jy)} &
{(\%)} &
{(Jy)} &
{(\%)} &
{(Jy)} &
{(\%)} &
{(Jy)} &
{(\%)} \\

\hline

\multicolumn{9}{c}{Water Fountains} \\
\hline
IRAS~15445$-$5449 & 0.652$\phantom{^{\diamond}}$ & 4.2 & 5.389$\phantom{^{\diamond}}$ & 5.0 & 13.670 & 6.1 & 45.580 & 6.0 \\
IRAS~15544$-$5332 & 2.974$\phantom{^{\diamond}}$ & 4.1 & 4.521$\phantom{^{\diamond}}$ & 5.1 & 6.735 & 6.1 & 8.649 & 6.1 \\
IRAS~16342$-$3814 & 1.540$\phantom{^{\diamond}}$ & 4.2 & 13.310$\phantom{^{\diamond}}$ & 5.0 & 43.650 & 6.1 & 125.600 & 6.0 \\
IRAS~16552$-$3050 & $\cdots$$\phantom{^{\diamond}}$ & $\cdots$ & $\cdots$$\phantom{^{\diamond}}$ & $\cdots$ & $\cdots$ & $\cdots$ & $\cdots$ & $\cdots$ \\
IRAS~18043$-$2116 & 0.485$\phantom{^{\diamond}}$ & 4.4 & 2.287$\phantom{^{\diamond}}$ & 5.6 & 4.796 & 6.1 & 4.328 & 6.3 \\
IRAS~18056$-$1514 & 10.020$\phantom{^{\diamond}}$ & 4.1 & 12.260$\phantom{^{\diamond}}$ & 5.0 & 14.140 & 6.1 & 10.520 & 6.1 \\
IRAS~18113$-$2503 & 0.691$\phantom{^{\diamond}}$ & 4.3 & 2.535$\phantom{^{\diamond}}$ & 5.7 & 5.475 & 6.1 & 9.635 & 6.1 \\
OH~12.8$-$0.9 & 8.132$\phantom{^{\diamond}}$ & 4.1 & 12.340$\phantom{^{\diamond}}$ & 5.0 & 18.930 & 6.1 & 13.660 & 6.0 \\
IRAS~18286$-$0959 & 29.680$\phantom{^{\diamond}}$ & 4.1 & 45.010$\phantom{^{\diamond}}$ & 5.0 & 61.540 & 6.1 & 33.410 & 6.0 \\
OH~16.3$-$3.0 & 11.860$\phantom{^{\diamond}}$ & 4.1 & 20.920$\phantom{^{\diamond}}$ & 5.0 & 24.440 & 6.1 & 27.080 & 6.0 \\
W~43A & 2.509$\phantom{^{\diamond}}$ & 4.1 & 23.870$\phantom{^{\diamond}}$ & 5.0 & 53.390 & 6.1 & 78.810 & 6.0 \\
IRAS~18455$+$0448 & 5.466$\phantom{^{\diamond}}$ & 4.1 & 9.325$\phantom{^{\diamond}}$ & 5.0 & 13.440 & 6.1 & 11.710 & 6.1 \\
IRAS~18460$-$0151 & $\cdots$$\phantom{^{\diamond}}$ & $\cdots$ & $\cdots$$\phantom{^{\diamond}}$ & $\cdots$ & $\cdots$ & $\cdots$ & $\cdots$ & $\cdots$ \\
IRAS~18596$+$0315 & 0.484$\phantom{^{\diamond}}$ & 4.4 & 2.940$\phantom{^{\diamond}}$ & 5.4 & 6.193 & 6.1 & 9.310 & 6.1 \\
IRAS~19134$+$2131 & 1.224$\phantom{^{\diamond}}$ & 4.2 & 5.058$\phantom{^{\diamond}}$ & 5.3 & 8.853 & 6.1 & 10.090 & 6.1 \\
IRAS~19190$+$1102 & 0.120$^{\diamond}$ & 8.0 & 1.175$^{\diamond}$ & 7.9 & 3.450 & 6.3 & 8.092 & 6.1 \\
IRAS~19356$+$0754 & $\cdots$$\phantom{^{\diamond}}$ & $\cdots$ & $\cdots$$\phantom{^{\diamond}}$ & $\cdots$ & $\cdots$ & $\cdots$ & $\cdots$ & $\cdots$ \\
\hline
\multicolumn{9}{c}{AGB Stars} \\
\hline
IRAS~14247$+$0454 & $\cdots$$\phantom{^{\diamond}}$ & $\cdots$ & $\cdots$$\phantom{^{\diamond}}$ & $\cdots$ & $\cdots$ & $\cdots$ & $\cdots$ & $\cdots$ \\
IRAS~18556$+$0811 & 72.930$\phantom{^{\diamond}}$ & 4.1 & 86.550$\phantom{^{\diamond}}$ & 5.0 & 65.560 & 6.1 & 72.790 & 6.0 \\
IRAS~19149$+$1638 & 24.130$\phantom{^{\diamond}}$ & 4.1 & 26.920$\phantom{^{\diamond}}$ & 5.0 & 20.180 & 6.1 & 22.040 & 6.0 \\
IRAS~19312$+$1130 & 9.494$\phantom{^{\diamond}}$ & 4.1 & 9.701$\phantom{^{\diamond}}$ & 5.0 & 7.005 & 6.1 & 7.889 & 6.1 \\
IRAS~19395$+$1827 & 4.098$\phantom{^{\diamond}}$ & 4.1 & 4.971$\phantom{^{\diamond}}$ & 5.2 & 3.641 & 6.3 & $\cdots$ & $\cdots$ \\
IRAS~19495$+$0835 & $\cdots$$\phantom{^{\diamond}}$ & $\cdots$ & $\cdots$$\phantom{^{\diamond}}$ & $\cdots$ & $\cdots$ & $\cdots$ & $\cdots$ & $\cdots$ \\
\hline
\multicolumn{9}{c}{Post-AGB Stars} \\
\hline
IRAS~07134$+$1005 & $\cdots$$\phantom{^{\diamond}}$ & $\cdots$ & $\cdots$$\phantom{^{\diamond}}$ & $\cdots$ & $\cdots$ & $\cdots$ & $\cdots$ & $\cdots$ \\
IRAS~19024$+$0044 & 0.679$\phantom{^{\diamond}}$ & 4.3 & 2.618$\phantom{^{\diamond}}$ & 5.6 & 8.407 & 6.1 & 32.610 & 6.0 \\
IRAS~22272$+$5435 & 25.070$\phantom{^{\diamond}}$ & 4.1 & 87.850$\phantom{^{\diamond}}$ & 5.0 & 95.380 & 6.1 & 186.600 & 6.0 \\
\hline
\end{tabular}
\\
\flushleft
$^{\diamond}${Unreliable flux value.}
\end{table*}


\begin{table*}
\caption{\textsl{AKARI} photometric data.}
\label{tab:akari}
\flushleft
\begin{tabular}{lrrrrrrrrrrrr}
\hline
{Object} &
{$9~\mu$m} &
{$\sigma_{9\mu{\rm m}}$} &
{$18~\mu$m} &
{$\sigma_{18\mu{\rm m}}$} &
{$65~\mu$m} &
{$\sigma_{65\mu{\rm m}}$} &
{$90~\mu$m} &
{$\sigma_{90\mu{\rm m}}$} &
{$140~\mu$m} &
{$\sigma_{140\mu{\rm m}}$} &
{$160~\mu$m} &
{$\sigma_{160\mu{\rm m}}$} \\

&
{(Jy)} &
{(Jy)} &
{(Jy)} &
{(Jy)} &
{(Jy)} &
{(Jy)} &
{(Jy)} &
{(Jy)} &
{(Jy)} &
{(Jy)} &
{(Jy)} &
{(Jy)} \\

\hline

\multicolumn{13}{c}{Water Fountains} \\
\hline
IRAS~15445$-$5449 & 0.655 & 0.01 & 28.080 & 0.07 & $\cdots$ & $\cdots$ & $\cdots$$\phantom{^{\diamond}}$ & $\cdots$ & $\cdots$ & $\cdots$ & $\cdots$ & $\cdots$ \\
IRAS~15544$-$5332 & 2.824 & 0.04 & 7.315 & 0.03 & $\cdots$ & $\cdots$ & $\cdots$$\phantom{^{\diamond}}$ & $\cdots$ & $\cdots$ & $\cdots$ & $\cdots$ & $\cdots$ \\
IRAS~16342$-$3814 & 1.776 & 0.03 & 89.450 & 2.09 & $\cdots$ & $\cdots$ & $\cdots$$\phantom{^{\diamond}}$ & $\cdots$ & $\cdots$ & $\cdots$ & $\cdots$ & $\cdots$ \\
IRAS~16552$-$3050 & 0.315 & 0.04 & 5.555 & 0.15 & 10.150 & 1.01 & 7.897$\phantom{^{\diamond}}$ & 0.53 & 4.157 & 1.36 & $\cdots$ & $\cdots$ \\
IRAS~18043$-$2116 & 0.316 & 0.02 & 4.347 & 0.04 & 11.210 & 3.46 & 9.279$\phantom{^{\diamond}}$ & 1.19 & $\cdots$ & $\cdots$ & $\cdots$ & $\cdots$ \\
IRAS~18056$-$1514 & 10.310 & 0.06 & 15.280 & 0.09 & 6.283 & 2.07 & 4.518$\phantom{^{\diamond}}$ & 0.87 & $\cdots$ & $\cdots$ & $\cdots$ & $\cdots$ \\
IRAS~18113$-$2503 & $\cdots$ & $\cdots$ & 6.715 & 0.05 & 16.190 & 5.25 & 14.930$\phantom{^{\diamond}}$ & 2.36 & $\cdots$ & $\cdots$ & $\cdots$ & $\cdots$ \\
OH~12.8$-$0.9 & 8.223 & 0.07 & 15.740 & 0.05 & $\cdots$ & $\cdots$ & 13.010$^{\diamond}$ & 2.64 & $\cdots$ & $\cdots$ & $\cdots$ & $\cdots$ \\
IRAS~18286$-$0959 & 21.980 & 0.39 & 45.840 & 0.19 & 51.580 & 4.57 & 25.830$\phantom{^{\diamond}}$ & 3.49 & $\cdots$ & $\cdots$ & $\cdots$ & $\cdots$ \\
OH~16.3$-$3.0 & 13.090 & 0.79 & 28.150 & 0.97 & 15.920 & 1.49 & 12.750$\phantom{^{\diamond}}$ & 0.50 & 8.112 & 3.35 & $\cdots$ & $\cdots$ \\
W~43A & 2.191 & 0.14 & $\cdots$ & $\cdots$ & $\cdots$ & $\cdots$ & $\cdots$$\phantom{^{\diamond}}$ & $\cdots$ & $\cdots$$\phantom{^{\diamond}}$ & $\cdots$ & $\cdots$ & $\cdots$ \\
IRAS~18455$+$0448 & 4.889 & 0.02 & 12.620 & 0.09 & 5.627 & 0.34 & 4.771$\phantom{^{\diamond}}$ & 0.16 & $\cdots$ & $\cdots$ & $\cdots$ & $\cdots$ \\
IRAS~18460$-$0151 & $\cdots$ & $\cdots$ & $\cdots$ & $\cdots$ & $\cdots$ & $\cdots$ & $\cdots$$\phantom{^{\diamond}}$ & $\cdots$ & $\cdots$ & $\cdots$ & $\cdots$ & $\cdots$ \\
IRAS~18596$+$0315 & 0.598 & 0.02 & 7.817 & 0.08 & $\cdots$ & $\cdots$ & $\cdots$$\phantom{^{\diamond}}$ & $\cdots$ & $\cdots$ & $\cdots$ & $\cdots$ & $\cdots$ \\
IRAS~19134$+$2131 & 1.116 & 0.01 & 9.708 & 0.15 & 7.776 & 0.61 & 6.266$\phantom{^{\diamond}}$ & 0.65 & 3.396 & 0.79 & $\cdots$ & $\cdots$ \\
IRAS~19190$+$1102 & 0.133 & 0.01 & 6.100 & 0.05 & 17.670 & 3.82 & 12.900$\phantom{^{\diamond}}$ & 2.45 & $\cdots$ & $\cdots$ & $\cdots$ & $\cdots$ \\
IRAS~19356$+$0754 & $\cdots$ & $\cdots$ & $\cdots$ & $\cdots$ & $\cdots$ & $\cdots$ & $\cdots$$\phantom{^{\diamond}}$ & $\cdots$ & $\cdots$ & $\cdots$ & $\cdots$ & $\cdots$ \\
\hline
\multicolumn{13}{c}{AGB Stars} \\
\hline
IRAS~14247$+$0454 & 111.500 & 20.00 & 80.670 & 0.27 & 9.944 & 0.58 & 7.189$\phantom{^{\diamond}}$ & 0.24 & $\cdots$ & $\cdots$ & $\cdots$ & $\cdots$ \\
IRAS~18556$+$0811 & 94.540 & 0.81 & 64.970 & 14.30 & $\cdots$ & $\cdots$ & $\cdots$$\phantom{^{\diamond}}$ & $\cdots$ & $\cdots$ & $\cdots$ & $\cdots$ & $\cdots$ \\
IRAS~19149$+$1638 & 25.910 & 0.01 & 16.310 & 0.03 & $\cdots$ & $\cdots$ & $\cdots$$\phantom{^{\diamond}}$ & $\cdots$ & $\cdots$ & $\cdots$ & $\cdots$ & $\cdots$ \\
IRAS~19312$+$1130 & 8.348 & 3.19 & 6.724 & 1.24 & $\cdots$ & $\cdots$ & 0.504$\phantom{^{\diamond}}$ & 0.08 & $\cdots$ & $\cdots$ & $\cdots$ & $\cdots$ \\
IRAS~19395$+$1827 & 7.761 & 1.77 & 7.281 & 1.06 & $\cdots$ & $\cdots$ & 1.002$\phantom{^{\diamond}}$ & 0.20 & $\cdots$ & $\cdots$ & $\cdots$ & $\cdots$ \\
IRAS~19495$+$0835 & 57.070 & 14.40 & 43.540 & 14.70 & 7.072 & 0.87 & 4.724$\phantom{^{\diamond}}$ & 0.69 & $\cdots$ & $\cdots$ & $\cdots$ & $\cdots$ \\
\hline
\multicolumn{13}{c}{Post-AGB Stars} \\
\hline
IRAS~07134$+$1005 & 8.909 & 0.06 & 66.250 & 1.15 & 51.260 & 2.64 & 26.750$\phantom{^{\diamond}}$ & 1.95 & 8.701 & 1.19 & $\cdots$ & $\cdots$ \\
IRAS~19024$+$0044 & 0.814 & 0.01 & 21.980 & 0.11 & $\cdots$ & $\cdots$ & $\cdots$$\phantom{^{\diamond}}$ & $\cdots$ & 6.178 & 0.59 & $\cdots$ & $\cdots$ \\
IRAS~22272$+$5435 & 31.000 & 0.21 & 148.800 & 1.32 & 83.770 & 1.85 & 36.870$\phantom{^{\diamond}}$ & 3.08 & 14.830 & 2.28 & 8.472 & 1.85 \\
\hline
\end{tabular}
\\
\flushleft
$^{\diamond}${Unreliable flux value.}
\end{table*}


\begin{table*}
\caption{\textsl{IRAC} and \textsl{MIPS} (24~$\mu$m) photometric data. The 
values are obtained by performing photometry on corresponding images, 
if applicable. Hence the data here are presented directly in terms of 
W~m$^{-2}$.}
\label{tab:spitzer}
\flushleft
\begin{tabular}{lrrrrrrrrrr}
\hline
{Object} &
{$3.6~\mu$m} &
{$\sigma_{3.6\mu{\rm m}}$} &
{$4.5~\mu$m} &
{$\sigma_{4.5\mu{\rm m}}$} &
{$5.8~\mu$m} &
{$\sigma_{5.8\mu{\rm m}}$} &
{$8.0~\mu$m} &
{$\sigma_{8.0\mu{\rm m}}$} &
{$24~\mu$m} &
{$\sigma_{24\mu{\rm m}}$} \\

&
\multicolumn{10}{c}{($10^{-13}$~W~m$^{-2}$)} \\

\hline

\multicolumn{11}{c}{Water Fountains} \\
\hline
IRAS~15445$-$5449 & 0.84 & 0.03 & 1.42 & 0.07 & 1.21 & 0.09 & 1.52 & 0.08 & 28.38$^{\ddagger}$ & $\cdots$\\
IRAS~15544$-$5332 & 8.33 & 0.84 & 10.88 & 0.34 & 2.01 & 0.22 & 10.81 & 0.18 & 8.87$^{\ddagger}$ & $\cdots$\\
IRAS~16342$-$3814 & $\cdots$ & $\cdots$ & $\cdots$ & $\cdots$ & $\cdots$ & $\cdots$ & $\cdots$ & $\cdots$ & $\cdots$$\phantom{^{\ddagger}}$ & $\cdots$\\
IRAS~16552$-$3050 & $\cdots$ & $\cdots$ & $\cdots$ & $\cdots$ & $\cdots$ & $\cdots$ & $\cdots$ & $\cdots$ & $\cdots$$\phantom{^{\ddagger}}$ & $\cdots$\\
IRAS~18043$-$2116 & 0.05 & 0.01 & 0.21 & 0.02 & 0.61 & 0.09 & 1.61 & 0.11 & 6.83$\phantom{^{\ddagger}}$ & 0.19\\
IRAS~18056$-$1514 & $\cdots$ & $\cdots$ & $\cdots$ & $\cdots$ & $\cdots$ & $\cdots$ & $\cdots$ & $\cdots$ & $\cdots$$\phantom{^{\ddagger}}$ & $\cdots$\\
IRAS~18113$-$2503 & $\cdots$ & $\cdots$ & $\cdots$ & $\cdots$ & $\cdots$ & $\cdots$ & $\cdots$ & $\cdots$ & $\cdots$$\phantom{^{\ddagger}}$ & $\cdots$\\
OH~12.8$-$0.9 & 6.22 & 0.23 & 10.86 & 0.28 & 39.87 & 0.90 & 29.42 & 0.89 & 12.33$^{\ddagger}$ & $\cdots$\\
IRAS~18286$-$0959 & $\cdots$ & $\cdots$ & 13.41 & 0.69 & 72.06 & 1.12 & 38.06 & 1.40 & 23.90$^{\ddagger}$ & $\cdots$\\
OH~16.3$-$3.0 & $\cdots$ & $\cdots$ & $\cdots$ & $\cdots$ & $\cdots$ & $\cdots$ & $\cdots$ & $\cdots$ & $\cdots$$\phantom{^{\ddagger}}$ & $\cdots$\\
W~43A & 0.07 & 0.01 & 0.08 & 0.01 & 1.27 & 0.04 & 9.54 & 0.44 & 135.20$\phantom{^{\ddagger}}$ & 26.00\\
IRAS~18455$+$0448 & $\cdots$ & $\cdots$ & $\cdots$ & $\cdots$ & $\cdots$ & $\cdots$ & $\cdots$ & $\cdots$ & $\cdots$$\phantom{^{\ddagger}}$ & $\cdots$\\
IRAS~18460$-$0151 & 0.20 & 0.01 & 2.35 & 0.22 & 13.99 & 0.33 & 19.29 & 0.11 & 20.04$^{\ddagger}$ & $\cdots$\\
IRAS~18596$+$0315 & 0.09 & 0.01 & 0.31 & 0.02 & 1.10 & 0.05 & 2.84 & 0.07 & 10.21$^{\ddagger}$ & $\cdots$\\
IRAS~19134$+$2131 & $\cdots$ & $\cdots$ & $\cdots$ & $\cdots$ & $\cdots$ & $\cdots$ & $\cdots$ & $\cdots$ & $\cdots$$\phantom{^{\ddagger}}$ & $\cdots$\\
IRAS~19190$+$1102 & $\cdots$ & $\cdots$ & $\cdots$ & $\cdots$ & $\cdots$ & $\cdots$ & $\cdots$ & $\cdots$ & $\cdots$$\phantom{^{\ddagger}}$ & $\cdots$\\
IRAS~19356$+$0754 & $\cdots$ & $\cdots$ & $\cdots$ & $\cdots$ & $\cdots$ & $\cdots$ & $\cdots$ & $\cdots$ & $\cdots$$\phantom{^{\ddagger}}$ & $\cdots$\\
\hline
\multicolumn{11}{c}{AGB Stars} \\
\hline
IRAS~14247$+$0454 & $\cdots$ & $\cdots$ & $\cdots$ & $\cdots$ & $\cdots$ & $\cdots$ & $\cdots$ & $\cdots$ & $\cdots$$\phantom{^{\ddagger}}$ & $\cdots$\\
IRAS~18556$+$0811 & $\cdots$ & $\cdots$ & $\cdots$ & $\cdots$ & $\cdots$ & $\cdots$ & $\cdots$ & $\cdots$ & $\cdots$$\phantom{^{\ddagger}}$ & $\cdots$\\
IRAS~19149$+$1638 & $\cdots$ & $\cdots$ & $\cdots$ & $\cdots$ & $\cdots$ & $\cdots$ & $\cdots$ & $\cdots$ & $\cdots$$\phantom{^{\ddagger}}$ & $\cdots$\\
IRAS~19312$+$1130 & $\cdots$ & $\cdots$ & $\cdots$ & $\cdots$ & $\cdots$ & $\cdots$ & $\cdots$ & $\cdots$ & $\cdots$$\phantom{^{\ddagger}}$ & $\cdots$\\
IRAS~19395$+$1827 & $\cdots$ & $\cdots$ & $\cdots$ & $\cdots$ & $\cdots$ & $\cdots$ & $\cdots$ & $\cdots$ & $\cdots$$\phantom{^{\ddagger}}$ & $\cdots$\\
IRAS~19495$+$0835 & $\cdots$ & $\cdots$ & $\cdots$ & $\cdots$ & $\cdots$ & $\cdots$ & $\cdots$ & $\cdots$ & $\cdots$$\phantom{^{\ddagger}}$ & $\cdots$\\
\hline
\multicolumn{11}{c}{Post-AGB Stars} \\
\hline
IRAS~07134$+$1005 & $\cdots$ & $\cdots$ & $\cdots$ & $\cdots$ & $\cdots$ & $\cdots$ & $\cdots$ & $\cdots$ & $\cdots$$\phantom{^{\ddagger}}$ & $\cdots$\\
IRAS~19024$+$0044 & $\cdots$ & $\cdots$ & $\cdots$ & $\cdots$ & $\cdots$ & $\cdots$ & $\cdots$ & $\cdots$ & 50.23$\phantom{^{\ddagger}}$ & 0.44\\
IRAS~22272$+$5435 & $\cdots$ & $\cdots$ & $\cdots$ & $\cdots$ & $\cdots$ & $\cdots$ & $\cdots$ & $\cdots$ & $\cdots$$\phantom{^{\ddagger}}$ & $\cdots$\\
\hline
\end{tabular}
\\
\flushleft
$^{\ddagger}${Lower band flux limit.}
\end{table*}


\begin{table*}
\caption{\textit{NMA} and \textit{JVLA} data for W~43A.}
\label{tab:submm}
\flushleft
\begin{tabular}{rrrrr}
\hline
{Frequency} &
{Flux} &
{$\sigma_{\rm Flux}$} &
{rms} &
{Beam size} \\

{(GHz)} &
{(mJy)} &
{(mJy)} &
{(mJy)} &
\\

\hline
\multicolumn{5}{c}{\textit{NMA}}\\
\hline
98.20 & 24.69$\phantom{^{\dagger}}$ & 4.84 & 1.72 & $3.8\arcsec\times 3.3\arcsec$ \\
110.20 & 31.94$\phantom{^{\dagger}}$ & 5.14 & 2.82 & $3.9\arcsec\times 2.9\arcsec$ \\
134.45 & 70.68$\phantom{^{\dagger}}$ & 23.10 & 11.02 & $3.6\arcsec\times 2.2\arcsec$\\
146.45 & 108.70$\phantom{^{\dagger}}$ & 33.00 & 10.51 & $2.8\arcsec\times 1.8\arcsec$\\
\hline
\multicolumn{5}{c}{\textit{JVLA}}\\
\hline
10.10 & 0.49$^{\dagger}$ & $\cdots$ & 0.16 & $4.48\arcsec\times 3.33\arcsec$ \\
24.20 & 0.24$\phantom{^{\dagger}}$ & 0.06 & 0.03 & $1.14\arcsec\times 0.93\arcsec$ \\
46.00 & 1.34$\phantom{^{\dagger}}$ & 0.08 & 0.04 & $0.54\arcsec\times 0.46\arcsec$ \\
\hline
\end{tabular}
\\
\flushleft
$^{\dagger}${3-$\sigma$ upper flux limit.}
\end{table*}



\bsp	
\label{lastpage}
\end{document}